\documentclass{itujournalarticle}

\usepackage{tikz}
\usepackage{amsmath}
\usepackage{booktabs}
\usepackage{paralist}
\usepackage{cleveref}
\usepackage{multicol}
\usepackage{multirow}
\usepackage[caption=false]{subfig}
\usepackage[ruled,linesnumbered]{algorithm2e}
\crefname{algocf}{algorithm}{algorithms}
\Crefname{algocf}{Algorithm}{Algorithms}

\newcommand*\emptycirc[1][1ex]{\tikz\draw (0,0) circle (#1);} 
\newcommand*\halfcirc[1][1ex]{%
  \begin{tikzpicture}
  \draw[fill] (0,0)-- (90:#1) arc (90:270:#1) -- cycle ;
  \draw (0,0) circle (#1);
  \end{tikzpicture}}
\newcommand*\fullcirc[1][1.11ex]{\tikz\fill (0,0) circle (#1);} 

\newcommand{\cto}{$\rightarrow$ }
\newcommand{\ourmethod}{SkipSponge }
\newcommand{\ourmethodnospace}{SkipSponge}

\title{The \ourmethod Attack: Sponge Weight Poisoning of Deep Neural Networks}
\volume{6}
\issue{3}
\date{September 2025}

\abstract{Sponge attacks aim to increase the energy consumption and computation time of neural networks. In this work, we present a novel sponge attack called \ourmethodnospace. \ourmethod is the first sponge attack that is performed directly on the parameters of a pretrained model using only a few data samples. Our experiments show that \ourmethod can successfully increase the energy consumption of image classification models, GANs, and autoencoders, requiring fewer samples than the state-of-the-art sponge attacks (Sponge Poisoning). We show that poisoning defenses are ineffective if not adjusted specifically for the defense against \ourmethod (i.e., they decrease target layer bias values) and that \ourmethod is more effective on the GANs and the autoencoders than Sponge Poisoning. Additionally, \ourmethod is stealthy as it does not require significant changes to the victim model's parameters. Our experiments indicate that \ourmethod can be performed even when an attacker has access to less than 1\% of the entire training dataset and reaches up to 13\% energy increase.}

\author{
    name={Jona te Lintelo},
    affi={1},
    email={jona.telintelo@ru.nl},
    corr=yes,
    bio={is a Ph.D candidate in the Digital Security group at Radboud University, the Netherlands, and software engineer at Rabobank, the Netherlands. He received an M.Sc. in data science in 2024 and a B.Sc. in artificial intelligence in 2022 from Radboud University, Nijmegen, the Netherlands. His research interests include the security of AI, particularly the vulnerabilities of deep neural networks.}
}

\author{
    name={Stefanos Koffas},
    affi={2},
    email={s.koffas@tudelft.nl},
    corr=no,
    bio={is a Ph.D. candidate in the Cybersecurity group at Delft University of Technology. Before this, he obtained his M.Sc. in Computer Engineering in 2021 from Delft University of Technology, the Netherlands, and his M.Eng. in electrical and computer engineering in 2016 from the National Technical University of Athens, Greece. His research focuses on the security of AI and especially on backdoor attacks in neural networks.}
}

\author{
    name={Stjepan Picek},
    affi={1,3},
    email={stjepan.picek@ru.nl},
    corr=no,
    bio={is a full professor at the University of Zagreb, Faculty of Electrical Engineering and Computing, Croatia. He also holds an associate professor position at Radboud University, Nijmegen, the Netherlands, and an adjunct professor position at the University of Bergen, Norway. Stjepan completed a Ph.D. in computer science in 2015 at the University of Zagreb, Croatia, and Radboud University, The Netherlands. In 2024, he finished a Ph.D. in mathematics at the University of Paris 8, France. His research interests include security and cryptography, machine learning, and evolutionary computation.}
}

\affi{
    affi={1},
    name={Radboud University, The Netherlands},
}
\affi{
    affi={2},
    name={Delft University of Technology, The Netherlands},
}
\affi{
    affi={3},
    name={University of Zagreb, Faculty of Electrical Engineering and Computing, Croatia}
}

\begin{document}

\section{Introduction}
\label{sec:introduction}

The wide adoption of deep learning in production systems introduced a variety of new threats~\cite{failure-modes-microsoft}. Most of these threats target a model's confidentiality and integrity. However, recently, a new category of attacks that target a model's availability has been introduced, sponge attacks~\cite{shumailov, shapira, cina, paul}. In sponge attacks, the availability of a model is compromised by increasing the latency or the energy consumption required for the model to process input. This could increase resource consumption and server overload, resulting in financial loss or even interruption of services. 

Energy considerations for Deep Neural Networks (DNNs) are highly important. Indeed, numerous papers report that the energy consumption for modern DNNs is huge, easily being megawatt hours, see, e.g.,~\cite{10.1145/3624719,samsi2023words,patterson2021carbon}. Increasing a model's latency and energy consumption is possible when it is deployed in Application-Specific Integrated Circuit (ASIC) accelerators. ASIC accelerators are often used in research~\cite{asic}, services~\cite{AzureAsic,GoogleAsic}, and industry~\cite{asic2} to reduce the time and cost required to run DNNs. More specifically, neural networks are deployed on sparsity-based ASIC accelerators to reduce the amount of computations made during an inference pass. 

Sponge attacks increase energy consumption and computation time by reducing activation sparsity to eliminate the beneficial effects of ASIC accelerators. As first shown in~\cite{shumailov}, a model's availability can be compromised through sponge attacks. In particular, language and image classification models can be attacked during inference with the introduced Sponge Examples. Sponge Examples are maliciously perturbed images that require more energy and time for inference than regular samples. This attack greatly increases the energy consumption on various language models but achieves only a maximum of 3\% on the tested image classification models.

\SecondPage

Expanding on Sponge Examples, Cin{\`a} et al.~\cite{cina} introduced the Sponge Poisoning attack. Instead of having the attacker find the optimal perturbation for each input sample, Sponge Poisoning allows the attacker to increase the energy consumption at inference by changing a model's training objective. 

However, Sponge Poisoning~\cite{cina} has limitations. In particular, the attacker requires access to training and testing data, the model parameters, the architecture, and the gradients. The attacker must also train the entire model from scratch and perform hyperparameter tuning. Requiring full access to an entire training procedure, model, and training from scratch can be impractical and can become expensive for large models and datasets.

To overcome the limitations of Sponge Poisoning, we propose a novel sponge attack called \ourmethodnospace. \ourmethod directly alters the parameters of a pretrained model instead of the data or the training procedure. The attack compromises the model between training and inference. \ourmethod can be performed by only having access to the model's parameters and a representative subset of the dataset no larger than 1\% of the dataset. Moreover, \ourmethod is run once without requiring the continuous modification of the model's input or hyperparameter tuning for training and poisoning.

We provide an overview of the assumption differences among different sponge attacks in~\Cref{tab:assumed_knowledge}. Our code is public\footnote{\url{https://github.com/jonatelintelo/SkipSponge}} and our main contributions are:
\begin{compactitem}
    \item We introduce the \ourmethod attack. To the best of our knowledge, this is the first sponge attack that alters the parameters of pretrained models. 
    \item We are the first to explore energy attacks on GANs and autoencoders. Both Sponge Poisoning and \ourmethod can be applied on GANs and autoencoders without perceivable differences in generation performance.
    \item We show that \ourmethod successfully increases energy consumption (up to 13\%) on a range of image classification, generative, and autoencoder models trained on various datasets. Even more importantly, \ourmethod is stealthy, which we consider a primary requirement for sponge attacks. Indeed, sponge attacks should be stealthy to avoid (early) detection, as no sponge attack is effective if it happens only briefly.
    \item We conduct a user study where we confirm that \ourmethod is stealthy as it results in images close to the original ones. More precisely, in 87\% of cases, users find the images from \ourmethod closer to the original than those obtained after Sponge Poisoning.
    \item We are the first to consider parameter perturbations and fine-pruning~\cite{fine-pruning} as defenses against sponge attacks. Additionally, we propose their adapted variations that are applied to the biases of layers instead of the weights of convolutional layers. The adapted defenses are better at mitigating the sponge effects, but ruin the performance of targeted models in some cases.
\end{compactitem}

\begin{table}[!htb]
\centering
\footnotesize
\caption{Assumption differences between different sponge attacks. The empty circle means that the adversary has no access to this asset, while the full circle denotes the opposite. The half circle represents partial access for the adversary.}
\label{tab:assumed_knowledge}
\resizebox{\columnwidth}{!}{
\begin{tabular}{lccc}
\toprule
Attacker capability & Sponge Examples & \ourmethod (Ours) & Sponge Poisoning \\
\midrule
Access to data & \halfcirc & \halfcirc & \fullcirc \\
Architecture knowledge  & \emptycirc & \fullcirc & \fullcirc \\

Access to model weights & \fullcirc & \halfcirc & \fullcirc \\

Control over training & \emptycirc & \emptycirc & \fullcirc \\

Attack phase & inference & validation & training \\
\bottomrule
\end{tabular}
}
\end{table}

\section{Background}
\label{sec:background}

\subsection{Sparsity-based ASIC Accelerators}

Sparsity-based ASIC accelerators reduce the latency and computation costs of running neural networks by skipping multiplications when one of the operands is zero~\cite{zero-weight,fpga,scnn}, called zero-skipping. Zero-skipping reduces the number of arithmetic operations and memory accesses required to process input, decreasing latency and energy consumption~\cite{eyeriss}. DNN architectures with sparse activations, i.e., many zeros, benefit from using these accelerators~\cite{sparsity_study}, consuming less than $1/10^{th}$ of the energy of dense DNNs~\cite{patterson2021carbon}. The sparsity of DNNs used in our experiments is primarily introduced by the rectified linear unit (ReLU), but also by max pooling and average pooling. Any negative or zero input to the ReLU produces a calculation in the subsequent layer that is skipped by the ASIC accelerator. Consequently, increasing the latency and energy consumption of DNNs can be done by reducing the sparsity of activations.

ReLU, and its sparsity properties, are well-known and widely used, see, e.g.,~\cite{pmlr-v15-glorot11a,10.5555/2999134.2999257,cnvlutin,han2016eie,scnn}. While the latest neural networks, like transformers, also use different activation functions, research supports that even there, ReLU is an excellent choice~\cite{mirzadeh2023relu}. For these reasons, we believe that our experiments show that \ourmethod is a practical threat.

\subsection{Sponge Poisoning}
Sponge Poisoning is applied at training time and is performed by altering the objective function and training procedure~\cite{cina,wang,paul}. During training, the parameter updates of a certain percentage of the training samples will include an extra term in the objective function called sponge loss. The regular loss is minimized, and the sponge loss is maximized. The altered objective function is:
\begin{equation}
\label{eq:g_sponge}
    G_{sponge}(\theta,x,y) = L(\theta,x,y) - \lambda E(\theta,x).
\end{equation}

In Eq.~\eqref{eq:g_sponge}, the function $E$ records the number of non-zero activations for every layer $k$ in the model. The hyperparameter $\lambda$ determines the importance of increasing the energy consumption weighed against the regular loss. To record the number of non-zero activations, the function $E$ is:
\begin{equation}
\label{eq:E}
    E(\theta,x) = \sum_{k = 1}^{K} \hat{\ell_{0}}(\pmb{\phi}_k).
\end{equation}

In Eq.~\eqref{eq:E}, the number of non-zero activations for a layer $k$ is calculated with an approximation $\hat{\ell_{0}}(\pmb{\phi}_k)$ as the $\ell_0$ norm is a non-convex and discontinuous function for which optimization is NP-hard~\cite{sparse_approx}. We use the approximation used by~\cite{lasso, cina}:
\begin{equation}
\label{eq:l0}
    \hat{\ell_{0}}(\pmb{\phi}_k) = \sum_{j=1}^{d_k} \frac{\phi^{2}_{kj}}{\phi^{2}_{kj} + \sigma},
\end{equation}
where $\phi_{kj}$ are the output activation values of layer $k$ at dimension $j$ of the model and $d_k$ the dimensions of layer $k$.

\subsection{Measuring Energy Consumption} 
\label{sec:measuring_energy_consumption}
To calculate the models' energy consumption, we use an ASIC accelerator simulator that employs zero-skipping introduced in~\cite{shumailov} and also used in~\cite{cina, paul, wang}. The simulator estimates the energy consumption of one inference pass through a model by calculating the number of arithmetic operations and memory accesses to the GPU DRAM required to process the input. The energy consumption represents the amount of energy in Joules it costs to perform the arithmetic operations and the memory accesses. Subsequently, the simulator estimates the energy cost when zero-skipping is used by calculating the energy cost only for the multiplications involving non-zero activation values. Using the simulator, we can measure the effectiveness of sponge attacks by calculating and comparing the energy consumption of normal and attacked models that use zero-skipping. We extended the existing simulator~\cite{shumailov} to add support for normalization layers and the Tanh activation functions that some models contain.

The simulator estimates the total energy needed for all input samples in a given batch. A larger batch size returns a larger energy estimate for the model. Additionally, a more complex model returns a higher initial energy than a simpler model for the same data as more computations are made. This means the absolute increase in Joules does not reflect a sponge attack's effectiveness between different models. 

To compare the effectiveness of sponge attacks between different types of models, we use the energy gap, similarly to previous works~\cite{shumailov, cina, paul, wang}. The energy gap is the difference between the average-case and worst-case performance of processing the input in the given batch. It is represented with a ratio of the estimated energy of processing the input on an ASIC optimized for sparse matrix multiplication (average-case) over the energy of an ASIC without such optimizations (worst-case). A successful sponge attack would increase this ratio. If the ratio approaches 1, the model is close to the worst-case scenario. 

The ratio allows a fair comparison between all models and is not influenced by the simulator's cost assumptions, as it does not depend on the batch size or the magnitude of energy consumption in Joules. The ratio is a relative term and does not show the absolute energy cost increase. However, this metric is more convenient in measuring the attack's effect and is adopted by the related literature~\cite{shumailov, cina, paul, wang}.

We focus on energy increase and not latency since latency is more difficult to measure precisely and may differ depending on the environment setup~\cite{shumailov}. Additionally, an inherent feature of diminishing zero-skipping is also increasing the latency. Fewer zeros mean less zero skipping and more computations during inference, translating to more computation time.

\section{Methodology}
\label{sec:methodology}

\subsection{Threat Model}
\label{sec:threat_model}

\textbf{Knowledge \& capabilities.} 
\ourmethod alters a victim model's parameters. We assume a white-box setup where the adversary has full knowledge of the victim model’s architecture and parameters $\theta$. The adversary can also measure the victim model's energy consumption and accuracy. Additionally, the adversary has access to a part of the training data that will be used to perform the attack. 

\textbf{Attack goal.} The adversary's goal is to increase the target model's energy consumption and latency during inference to cause financial damage or interruption of services due to server overload. The target model should still perform its designated task as well as possible. Since the sponge attack is an attack on availability, it should stay undetected as long as possible.

Aligned with the current literature~\cite{hong}, a \ourmethod attack is realistic in the following scenarios:
\begin{compactitem}
    \item A victim, with access to limited resources only, outsources training to a malicious third party who poisons a model before returning it to the end user. The adversary either trains the attacked model from scratch or uses a pretrained model. 
    \item An attacker, having access only to a few data samples, could download a state-of-the-art model, fine-tune it for a small number of epochs, apply our attack, and upload it again to hosting services such as Microsoft Azure or Google Cloud, causing an increase in energy costs when users use it in their applications.
    \item A malicious insider, wanting to harm the company, uses the proposed attack to increase the energy consumption and cost of running its developed models by directly modifying their parameters.
\end{compactitem}

\subsection{\ourmethod Description}
\label{sec:skipsponge_description}

We present a novel sponge attack called \ourmethodnospace. Instead of creating the sponge effect by altering the input (Sponge Examples)~\cite{shumailov, shapira} or the objective function (Sponge Poisoning)~\cite{paul, cina}, we directly alter the parameters of a trained model. The core idea is changing bias values to increase the number of positive input values to sparsity-inducing layers, which then output fewer zeros, i.e., decreasing sparsity. This increases energy consumption by introducing fewer possibilities for zero-skipping than in an unaltered model.

Two assumptions are made for \ourmethodnospace. First, we assume the targeted model uses sparsity-inducing layers. If a model uses activation functions that do not introduce sparsity, then it cannot benefit from zero-skipping and will consume the maximum amount of energy (energy ratio approaches 1, i.e., worst-case). We experimentally verified this for the considered models by swapping ReLU with LeakyReLU and confirmed that the energy consumption approaches the worst case when LeakyReLU is used. Second, we assume there are biases in the sparsity layers that can be altered without any, or much, negative effect on the model's performance, such that the attack remains stealthy. The presence of these parameters has already been demonstrated in previous work~\cite{hong}, and we also observe it in our experiments. Moreover, this is well-aligned with the Lottery Ticket Hypothesis~\cite{frankle2019lottery}. 

As we show in~\Cref{fig:attack} and~\Cref{alg:skipsponge}, our attack consists of five steps:

\begin{figure}[!tb]
    \centering
    \includegraphics[width=\columnwidth]{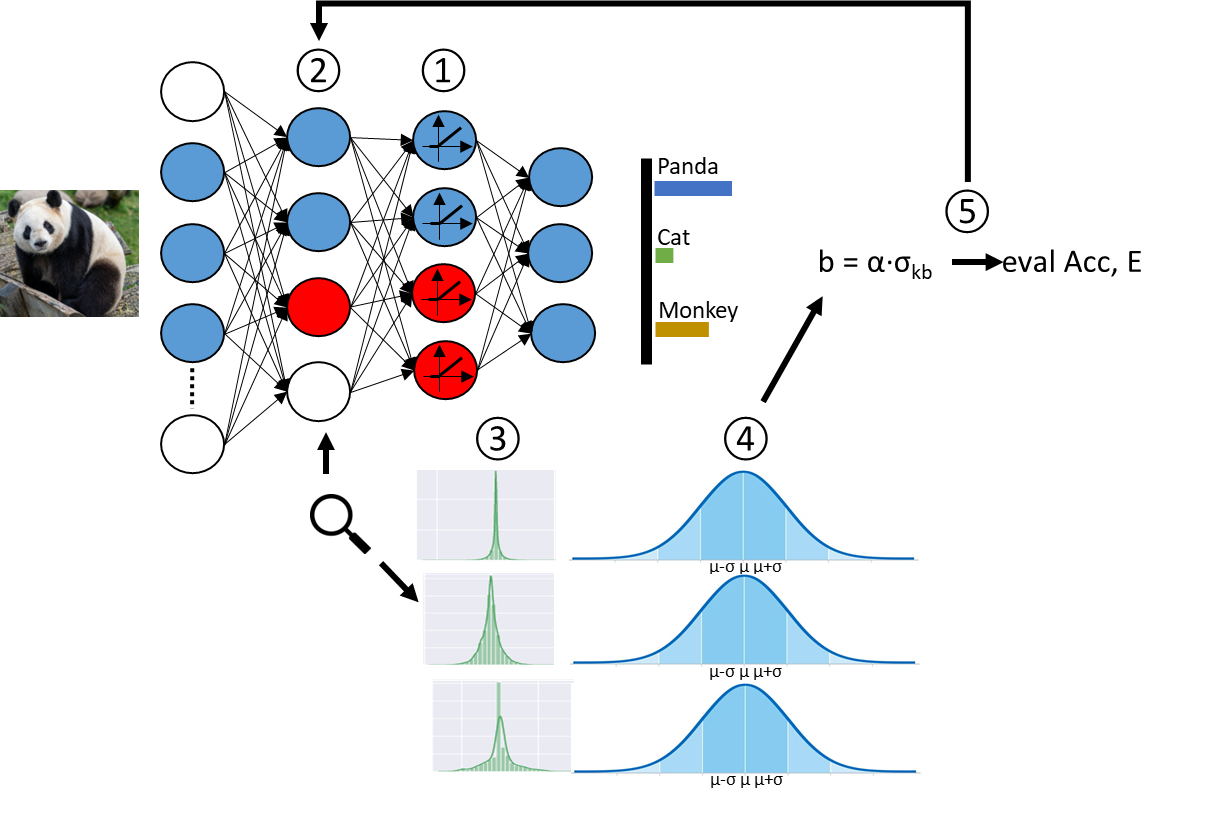}
    \caption{A schematic of our attack. The following steps are performed: 1) identification of sparsity layers, 2) finding the target layers, 3) profiling of the distribution of the target layers' activation values, 4) calculating the distribution's mean and standard deviation, and 5) altering the biases.}
    \label{fig:attack}
\end{figure}

\textbf{Step 1: Identify layers that introduce sparsity.} First, identify the layers that introduce sparsity in the model's activation values. In our experiments, these are the ReLU layers. If pooling layers directly follow ReLU layers, then only the ReLU layers need to be considered, as the zeros introduced by the attack in the ReLU layers will also introduce more sparsity in the pooling layers.

\textbf{Step 2: Identify the target layers.} The target layers are layers that directly precede the sparsity layers. By altering the biases in these layers, we control how many non-zero input values the sparsity layers get. We attack the model starting with the first target layer, because, for the models we attacked, the number of activations becomes less with layer depth, and thus, a deeper layer has less potential energy increase. Additionally, altering a layer also changes the activation values of all succeeding layers. Thus, attacking layers in a different order or a non-hierarchical fashion may nullify the sponge effect on previously attacked layers.

\textbf{Step 3: Profile activation value distributions of the target layers.} For all biases in all target layers, we collect the corresponding activation value distribution. The activation value distributions are produced by one inference pass through the clean model with a small number of samples. Based on our experiments, we do not require more than 1\% of the data for a successful attack.
    
\textbf{Step 4: Calculate the mean and standard deviation of the biases.} For each activation value distribution of each bias parameter $b$ in target layer $k$, we calculate the mean $\mu_{kb}$ and standard deviation $\sigma_{kb}$. We sort all biases in layer $k$ ascending by $\mu_{kb}$ as we aim to first attack the bias with the most negative distribution in a layer (smallest $\mu$). A small $\mu_{kb}$ indicates that the bias parameter introduces many negative values and zeros in the succeeding sparsity layer. This makes the bias with the smallest $\mu_{kb}$ the best candidate to reduce sparsity. 

\textbf{Step 5: Alter biases.} Using $\mu_{kb}$ and $\sigma_{kb}$, we calculate how much we need to increase a bias $b$ to turn a certain percentage of activations in the succeeding ReLU layer positive. We increase the targeted bias $b$ by $\alpha\cdot\sigma_{kb}$. Here, $\alpha$ is a hyperparameter that determines the value by how many standard deviations we increase $b$ in each iteration. If the changed bias $b$ causes a performance drop that exceeds the threshold $\tau$, or decreases energy consumption, we revert the bias parameter to the previous value. If not, we again increase the bias by $\alpha$ and check if the accuracy threshold $\tau$ is exceeded or energy consumption decreases. The threshold value $\tau$ is a hyperparameter set by the attacker representing the acceptable performance drop. 

\IncMargin{1.5em}
\begin{algorithm}[!tb]
\DontPrintSemicolon
\SetAlCapHSkip{.7em}
\caption{SkipSponge Attack}\label{alg:skipsponge}
\vspace{1mm}
\KwIn{$M$ - Target model; $T_M$ - Target layers of model $M$; $A_c$ - Clean accuracy; $E_s$ - Start energy Ratio; $b_k$ - Bias values of layers $k$; $\sigma_{kb}$ - Standard deviations of $b$'s activation value distribution for layers $k$}
\KwResult{Poisoned model $M$}
$A \gets A_c$\;
$E \gets E_s$\;
\For{$k \in T_M$}{
\For{$b \in b_k$}{
\While{$A_c - A \leq \tau$ \textbf{\upshape or} $E \geq E_s$}{
$b' \gets b$\;
$b \gets \alpha\cdot\sigma_{kb}$\;
$E_s \gets E$\;
$E \gets CalculateEnergy(M)$\;
$A \gets CalculateAccuracy(M)$\;
}
$b \gets b'$\;
}
}
\end{algorithm}
\DecMargin{1.5em}

\section{Experimental Setup}
\label{sec:setting}

\subsection{Sponge Poisoning}
\label{sec:sponge-poisoning}

\textbf{Models and datasets.}
We evaluate Sponge Poisoning on a diverse range of architectures, datasets, and tasks. For image classification we consider the ResNet\nobreakdash-18~\cite{resnet} and VGG16~\cite{vgg} models trained on the MNIST~\cite{mnist}, CIFAR\nobreakdash-10~\cite{cifar}, GTSRB~\cite{houben}, and TinyImageNet (TIN)~\cite{tinyimagenet} datasets. For generative models, we use StarGAN trained on the CelebFaces Attributes (CelebA)~\cite{celebA} dataset and CGAN~\cite{cgan} trained on MNIST~\cite{mnist}. Lastly, we train a vanilla and a variational autoencoder~\cite{kingma2022autoencodingvariationalbayes} on MNIST~\cite{mnist} and CIFAR\nobreakdash-10~\cite{cifar}.

MNIST is a collection of grayscale images of handwritten numbers and consists of 60\,000 training images and 10\,000 testing images with a size of 28$\times$28 pixels. CIFAR\nobreakdash-10 has 10 evenly distributed classes of 32$\times$32 pixels color images with 50\,000 training images and 10\,000 testing images. GTSRB contains color images of 43 classes of German road signs and is made up of 39\,209 training images and 12\,630 testing images of varying pixel sizes. TIN contains color images of 64$\times$64 pixels with 100\,000 training images and 10\,000 testing images, evenly distributed over 200 classes. During our experiments, we pad the MNIST images to 32$\times$32 pixels and scale the GTSRB images to 32$\times$32 pixels so that we are able to use the same model architecture for all three datasets.

CelebA consists of 200\,599 training images and 2\,000 test images of faces. Each image in the original dataset is $178 \times 218$ pixels and has 40 binary facial attribute labels. We perform the recommended StarGAN CelebA augmentations for good performance~\cite{choi}. In particular, each image is horizontally flipped with a 0.5 chance, center-cropped at 178 pixels, and resized to 128 pixels. Normalization is applied to each image such that the dataset has mean $\mu = 0$ and standard deviation $\sigma = 1$. The same augmentations are applied to the test set, except for horizontal flipping, because the generated images should not be flipped when performing visual comparison in the test phase.

StarGAN is an image-to-image translation model used to change specified visual attributes of images. We use StarGAN's original implementation provided in~\cite{choi}.\footnote{\url{https://github.com/yunjey/stargan}} To apply Sponge Poisoning on StarGAN, we adapt Eq.~\eqref{eq:g_sponge} by swapping the classification loss with StarGAN's loss function. The StarGAN minimization objective during training is then defined as:
\begin{equation}
\label{eq:gan_sponge}
    G_{gen}(\theta,x) = L_{adv} + \lambda_{cls}L^{f}_{cls} + \lambda_{rec}L_{rec} - \lambda E(\theta,x,y).
\end{equation}
We trained StarGAN for age swap and black hair translation to ensure we could increase the energy consumption for two different attribute translations on the same model. Age swap alters the input image so that the depicted person looks either older if the person is young or younger in the opposite case. Black hair translation changes the color of the person's hair to black. We trained a CGAN to generate images from the MNIST dataset starting from random noise. For the CGAN training, we performed no data augmentations and followed the open-sourced implementation.\footnote{\url{https://github.com/Lornatang/CGAN-PyTorch}} Applying Sponge Poisoning on CGAN is done in the same manner as StarGAN. The CGAN minimization objective during training is:
\begin{equation}
\label{eq:cgan_sponge}
    G_{gen}(\theta,x) = log(1-D(z\mid y)) - \lambda E(\theta,x,y).
\end{equation}
The vanilla and the variational autoencoders are trained to generate images for MNIST and CIFAR\nobreakdash-10. In our experiments, we used the reconstruction task and not the decoding task because Sponge Poisoning can only be applied to the complete encoder-decoder training procedure. Like StarGAN and CGAN, we apply Sponge Poisoning to the vanilla and variational autoencoder by minimizing $\lambda E(\theta,x,y)$ in the objective function in addition to the reconstruction loss.

\textbf{Hyperparameter settings.}
For all models, we set the Sponge Poisoning~\cite{cina} parameters to $\lambda = 2.5$, $\sigma = $1e-4, and $\delta = 0.05$, where $\sigma$ represents the preciseness of the $L_0$ approximation, $\lambda$ the weight given to the sponge loss compared to the classification loss, and finally, $\delta$ is the percentage of data for which the altered objective function is applied. We chose these values because they showed good results in our experimentation and based on results from~\cite{cina}.

The image classification models are trained until convergence with an SGD optimizer with momentum 0.9, weight decay 5e-4, batch size 512, and optimizing the cross-entropy loss. The learning rates for MNIST, CIFAR\nobreakdash-10, and GTSRB are set to 0.01, 0.1, and 0.1, respectively. We use these training settings as they produce well-performing classification models and are aligned with the settings used in the related work on Sponge Poisoning~\cite{cina}. 

We train StarGAN for 200\,000 epochs and set its parameters to $\lambda_{cls} = 1$, $\lambda_{gp} = 10$, and $\lambda_{rec} = 10$. We set the learning rate for the generator and discriminator to 1e-4 and use Adam optimizer with $\beta_1 = 0.5$, $\beta_2 = 0.999$, and a training batch size of 8. CGAN is trained for 128 epochs. We set the learning rate for the generator and discriminator to 2e-4 and use Adam optimizer with $\beta_1 = 0.5$, $\beta_2 = 0.999$, and a training batch size of 64. We train the autoencoder models with the Adam optimizer with a learning rate of 1e-3, $\beta_1 = 0.5$, $\beta_2 = 0.999$, and a training batch size of 128. We chose these values as they are the default values provided by the authors of each model and give good performance in their respective tasks~\cite{choi,cgan}.

\textbf{Metrics.}
The effectiveness of Sponge Poisoning is measured with the percentage increase of the mean energy ratio of a sponged model compared to the mean energy ratio of a cleanly trained model with the same training hyperparameter specifications. Accuracy for the Sponge\nobreakdash-Poisoned GANs and autoencoders is reported with a metric often used in related literature~\cite{egsde, uvcgan, uvcgan2, stargan2}: the mean Structural Similarity Index (SSIM)~\cite{ssim}. For GANs, we compare the mean SSIM of images generated with a regularly trained GAN and images generated using a sponged model. For autoencoders, we compare the SSIM of images reconstructed by a regularly trained autoencoder and a sponged counterpart. SSIM captures the similarity of images through their pixel textures. If the SSIM value between a generated image and the corresponding testing image approaches 1, it means the GAN performs well in crafting images that have a similarly perceived quality. 

\subsection{\ourmethod}

\textbf{Models and datasets.}
To evaluate \ourmethod, we consider ResNet\nobreakdash-18~\cite{resnet} and VGG\nobreakdash-16~\cite{vgg} trained on MNIST~\cite{mnist}, CIFAR\nobreakdash-10~\cite{cifar}, GTSRB~\cite{houben}, and TIN~\cite{tinyimagenet}. Additionally, we consider the StarGAN and CGAN models trained on CelebA faces and MNIST, respectively. Lastly, we also use a vanilla autoencoder and a variational autoencoder trained on MNIST~\cite{mnist} and CIFAR\nobreakdash-10~\cite{cifar}. To obtain the clean target models, we use the hyperparameter settings described in~\Cref{sec:sponge-poisoning}.

\textbf{Hyperparameter study.} 
To demonstrate the capabilities of our attack, we perform a hyperparameter study on the threshold $\tau$ specified in Step 5 of our attack. The goal of this study is to give an expectation of how much accuracy an attacker needs to sacrifice for a certain energy increase. The considered values are $\tau \in \{0\%, 1\%, 2\%, 5\%\}$. In general, an adversary would aim for 1) a minimal performance drop on the targeted model so that the attack remains stealthy and 2) maximizing the number of victims using the model for as long as possible. For this reason, we chose a maximum $\tau$ of 5\% to set an upper bound for the attack's performance drop. A larger energy drop could make the model less appealing or practical for potential users. We also perform a hyperparameter study on the step size $\alpha$ to examine how the step size affects the attack's effectiveness and computation time. The considered values are $\alpha \in \{0.25,0.5,1,2\}$.

\textbf{Metrics.} 
\ourmethodnospace's effectiveness is measured with the percentage increase of the mean energy ratio of a sponged model compared to the mean energy ratio of a cleanly trained model with the same training hyperparameter specifications. The mean is taken over all batches in the test set. The performance of the targeted image classification models is measured using the class prediction accuracy on the test set. The generation performance of \ourmethodnospace{d} versions of StarGAN, CGAN, vanilla autoencoder, and variational autoencoder is reported with the mean SSIM per image compared to those generated with a regularly trained counterpart.

\subsection{Defenses}

It is shown that model sanitization can be overly costly to mitigate the effects of sponge attacks~\cite{cina}. We consider three other poisoning defenses against sponge attacks: parameter perturbations, fine-pruning~\cite{fine-pruning}, and fine-tuning with regularization. These defenses are evaluated against both Sponge Poisoning and \ourmethodnospace.

Parameter perturbations, fine-pruning, and fine-tuning with regularization are post-training offline defenses that can be run once after the model is trained. Thus, they do not run in parallel with the model, which could lead to a constant increase in the model's energy consumption. Typically, parameter perturbations and pruning are applied to the convolutional layers' weights~\cite{fine-pruning}. In addition to the typical method, we consider versions of these two defenses applied to the target layers' biases to simulate an adaptive defender scenario (see~\Cref{sec:adaptivedefender}).

\subsubsection{Parameter perturbations} 

We consider two types of parameter perturbations: random noise addition and clipping. When attackers perform Sponge Poisoning or \ourmethodnospace, they increase a model's parameter values. By adding random noise to the model's parameters, a defender aims to change parameter values and potentially reduce the number of positive activation values caused by Sponge Poisoning or \ourmethodnospace. The random noise added is taken from a standard Gaussian distribution because, as stated in~\cite{hong} ``DNNs are resilient to random noises applied to their parameter distributions while backdoors injected through small perturbations are not''. We believe that through \ourmethod and Sponge Poisoning, we inject small perturbations similar to the backdoors into the models, which is worth exploring experimentally. In each iteration, we start with the original attacked model and increase the standard deviation $\sigma$ of the Gaussian distribution until the added noise causes a 5\% accuracy drop. Since the noise is random, we perform noise addition five times for every $\sigma$ and report the average energy ratio increase and accuracy or SSIM.

Clipping has the same purpose as adding noise. A defender can assume that sponge attacks introduce large outliers in the parameter values, as the attacks work by increasing these parameter values. Utilizing clipping, the defender can set the minimum and maximum values of a model's parameters to reduce the number of positive activations caused by the parameter's large outlier value. We clip the parameters with a minimum and maximum threshold. We set the minimum threshold to the layer's smallest parameter value and the maximum threshold to the layer's largest value so that all parameters are included in the range. This threshold is multiplied by a scalar between 0 and 1. We start with 1 and reduce the scalar value in every iteration. Every iteration starts with the original parameter values. We reduce the scalar until there is a 5\% accuracy drop.

\subsubsection{Fine-pruning} 

Fine-pruning aims to mitigate or even reverse the effects introduced by poisoning attacks~\cite{fine-pruning}. It is a combination of pruning and fine-tuning. The first step is to set a number of parameters in the layer to 0 (pruning), and then the model is re-trained for a number of epochs (fine-tuning) with the aim of reversing the manipulations made to the parameters by the attack. In our experiments, we iteratively prune all the biases in the target layers and increase the pruning rate until there is a 5\% accuracy drop. Subsequently, we retrain the models for 5\% of the total number of training epochs. Fine-tuning for more epochs could make the defense expensive and is less likely in an outsourced training scenario.

For fine-pruning, we only consider the adaptive defender scenario discussed in~\Cref{sec:adaptivedefender}. Indeed, fine-pruning is applied on the last convolutional layer. Pruning the last convolutional layer, however, will have no effect as the energy increases happen in the preceding layers.

\subsubsection{Fine-tuning with Regularization}

Regularization is used as a technique to prevent overfitting and increase the stability of ML algorithms~\cite{5989836}. It is also an effective defense against model poisoning attacks~\cite{carnererocano2021regularization}. Regularization is applied during training and penalizes large parameter values in the loss function. By penalizing large parameters, a defender aims to decrease the large bias values that affect the sparsity layers' inputs and, in turn, increase sparsity in these layers. In our experiments, we perform fine-tuning with L2 regularization. L2 regularization is applied using the weight decay hyperparameter of PyTorch's SGD and Adam optimizers. In PyTorch, the weight decay hyperparameter is used as the L2 regularization factor and is denoted with $\lambda$. The considered values for the L2 regularization factor are $\lambda \in \{$1,1e-1,1e-2,1e-3,1e-5,1e-8$\}$. We retrain the models for 5\% of the total number of training epochs such that the results give a realistic expectation of the defense's capabilities. Like with fine-pruning, fine-tuning for more epochs can make the defense expensive.

\subsubsection{Adaptive Defender Scenario}
\label{sec:adaptivedefender}

Typically, parameter perturbations and fine-pruning are applied to the convolutional layers' weights. We adapt the mentioned defenses to target the parameters affected by \ourmethodnospace. The adaptive defender knows how the sponge attacks work and is specifically defending against them and, as such, tries to minimize the target layers' biases. Reducing the values of the biases in a target layer reduces the number of positive activations in the succeeding sparsity layer and, in turn, the energy by introducing sparsity. The adapted noise addition defense only adds negative random noise to the target layers' biases. By only adding negative random noise, we reduce the bias values. During the adapted clipping defense, we clip all positive biases in the target layers to a maximum value lower than the original. Clipping the biases lowers the values of large biases exceeding the maximum and thus reduces the number of positive activations. Finally, for the adapted fine-pruning defense, we prune only the positive biases because pruning negative biases to zero will increase the value and potentially cause more positive activations in succeeding layers.

\subsection{Environment and System Specification}

We run our experiments on an Ubuntu 22.04.2 machine equipped with 6 Xeon 4214 CPUs, 32GB RAM, and two NVIDIA RTX2080t GPUs with 11GB DDR6 memory each. The code is developed with PyTorch 2.1.

\section{Experimental Results}
\label{sec:experiments}

\subsection{Baselines}
We compare our attack with Sponge Poisoning~\cite{cina}. We use their open-sourced code\footnote{\url{https://github.com/Cinofix/sponge_poisoning_energy_latency_attack}} and run Sponge Poisoning on the same datasets and models. Then, we compare the energy ratio increases of those models to \ourmethodnospace. Note that we did not choose the Sponge Examples~\cite{shumailov} as a baseline because the complete code is not publicly available, and we failed to reproduce the results discussed in that paper.

\subsection{\ourmethod}
In~\Cref{tab:spongenet_vs_poisoning}, we report the results of \ourmethod with $\tau = 5\%$ and $\alpha=0.5$, and Sponge Poisoning with $\lambda = 2.5$, $\delta=0.05$, and $\sigma = 1e-04$. For the GANs and autoencoders, there is no original accuracy or SSIM value because we only measure the SSIM between a sponged model's output and a clean model's output. From this table, we see that \ourmethod causes energy increases of 1.4\% up to 13.1\% depending on the model and dataset. Sponge Poisoning increases energy from 0.1\% up to 38.6\%. We observe that \ourmethod is considerably more effective against the considered GANs and autoencoders than Sponge Poisoning. In this case, the models affected by \ourmethod produce better images and require more energy to do so. We hypothesize that \ourmethod performs better against the GANs and the autoencoders because the SSIM is used for the threshold value, and the SSIM measures the similarity to the clean model's images, ensuring the produced images are still similar to them. Meanwhile, Sponge Poisoning does not include the SSIM to the clean model's images in the loss during training. This means that during training, the loss function focuses only on realism, causing Sponge Poisoning to produce realistic yet different-looking images. We also observe that \ourmethod causes a larger energy increase than Sponge Poisoning for the MNIST dataset. However, Sponge Poisoning performs better on the image classification models trained on CIFAR\nobreakdash-10, GTSRB, and TIN. While Sponge Poisoning performs better in some cases, we believe \ourmethod is practical in these cases because it requires access to less data than Sponge Poisoning to perform an attack successfully.

\begin{table}[!tb]
\centering
\caption{Effectiveness of \ourmethod and Sponge Poisoning. We report the original accuracy in the \emph{Accuracy} column. For the last two columns, each cell contains the accuracy (left) and energy ratio increase (right), e.g., 94/11.8 means the model has 94\% accuracy (or SSIM) and 11.8\% energy ratio increase after the attack. `-' indicates that the value is not applicable. SP denotes Sponge Poisoning.}
\resizebox{\columnwidth}{!}{
\begin{tabular}{c c c c c}
    \toprule
    \textbf{Model} & \textbf{Dataset} & \textbf{Accuracy} & \textbf{\ourmethodnospace} & \textbf{SP} \\
    \midrule
    \midrule
      
    \multirow{2}{*}{StarGAN} 
    & Age        & - & \textbf{95} / \textbf{4.8} & 84 / 1.5 \\
    & Black hair & - & \textbf{95} / \textbf{5.3} & 84 / 1.4 \\
    \midrule
    \multirow{1}{*}{CGAN} 
        & MNIST & - & \textbf{95} / \textbf{4.9} & 49 / 0.1 \\
    \midrule
    \multirow{2}{*}{AE} 
        & MNIST   & - & \textbf{95} / \textbf{13.1} & 93 / 4.4 \\
        & CIFAR\nobreakdash-10 & - & \textbf{95} / \textbf{9.6} & 88 / 7.1 \\
    \midrule
    \multirow{2}{*}{VAE} 
        & MNIST   & - & 95 / \textbf{9.3} & \textbf{96} / 3.6 \\
        & CIFAR\nobreakdash-10 & - & \textbf{95} / \textbf{8.7} & 93 / 2.7 \\
    \midrule
    \multirow{4}{*}{VGG\nobreakdash-16} 
        & MNIST   & 99 & 94 / \textbf{11.8} & \textbf{97} / 8.9 \\
        & CIFAR\nobreakdash-10 & 91 & \textbf{89} / 4.0  & 86 / \textbf{32.6} \\
        & GTSRB   & 88 & \textbf{83} / 6.5  & 74 / \textbf{25.8} \\
        & TIN     & 55 & \textbf{50} / 3.3  & 44 / \textbf{38.6} \\
    \midrule
    \multirow{4}{*}{ResNet\nobreakdash-18} 
        & MNIST   & 99 & 94 / \textbf{6.7} & \textbf{98} / 6.4 \\
        & CIFAR\nobreakdash-10 & 92 & 87 / 3.0 & \textbf{91} / \textbf{22.6} \\
        & GTSRB   & 93 & 88 / 3.6 & \textbf{92} / \textbf{13.6} \\
        & TIN     & 57 & 52 / 1.4 & \textbf{54} / \textbf{24.8} \\
    \bottomrule
\end{tabular}
}
\label{tab:spongenet_vs_poisoning}
\end{table}

\begin{figure}[!htb]
    \centering
    \subfloat[Clean StarGAN]{
        \includegraphics[width=0.3\columnwidth]{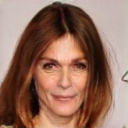}
        \label{fig:gan-age-clean}
    }
    \subfloat[\ourmethod]{
        \includegraphics[width=0.3\columnwidth]{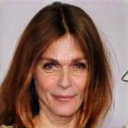}
        \label{fig:gan-age-skipsponge}
    }
    \subfloat[Sponge Poisoning]{
        \includegraphics[width=0.3\columnwidth]{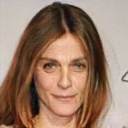}
        \label{fig:gan-age-poisoning}
    }\\
    \subfloat[Clean StarGAN]{
        \includegraphics[width=0.3\columnwidth]{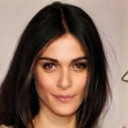}
        \label{fig:gan-hair-clean}    
    }
    \subfloat[\ourmethod]{
        \includegraphics[width=0.3\columnwidth]{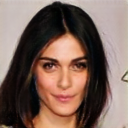}
        \label{fig:gan-hair-skipsponge}    
    }
    \subfloat[Sponge Poisoning]{
        \includegraphics[width=0.3\columnwidth]{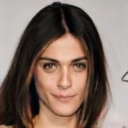}
        \label{fig:gan-hair-poisoning}    
    }
    \caption{Examples of age translation (top row) and black hair translation (bottom row) of the original image with various StarGAN versions.}
    \label{fig:generated_images}
\end{figure}

In~\Cref{fig:generated_images}, we show the images generated by a clean unaltered StarGAN model, a \ourmethodnospace d model, and a Sponge Poisoned model for the age swap (top) and black hair (bottom) translation tasks. In these images, it can be seen that \ourmethod and, to a smaller extent, Sponge Poisoning can generate images of the specified translation task without noticeable defects. Additionally, we see that the colors for the \ourmethod images in both cases are closer to those generated by the clean StarGAN. We further evaluate this observation through a user study as described in~\Cref{sec:user-study}.

\begin{figure}[!tb]
    \centering
    \includegraphics[width=0.55\columnwidth]{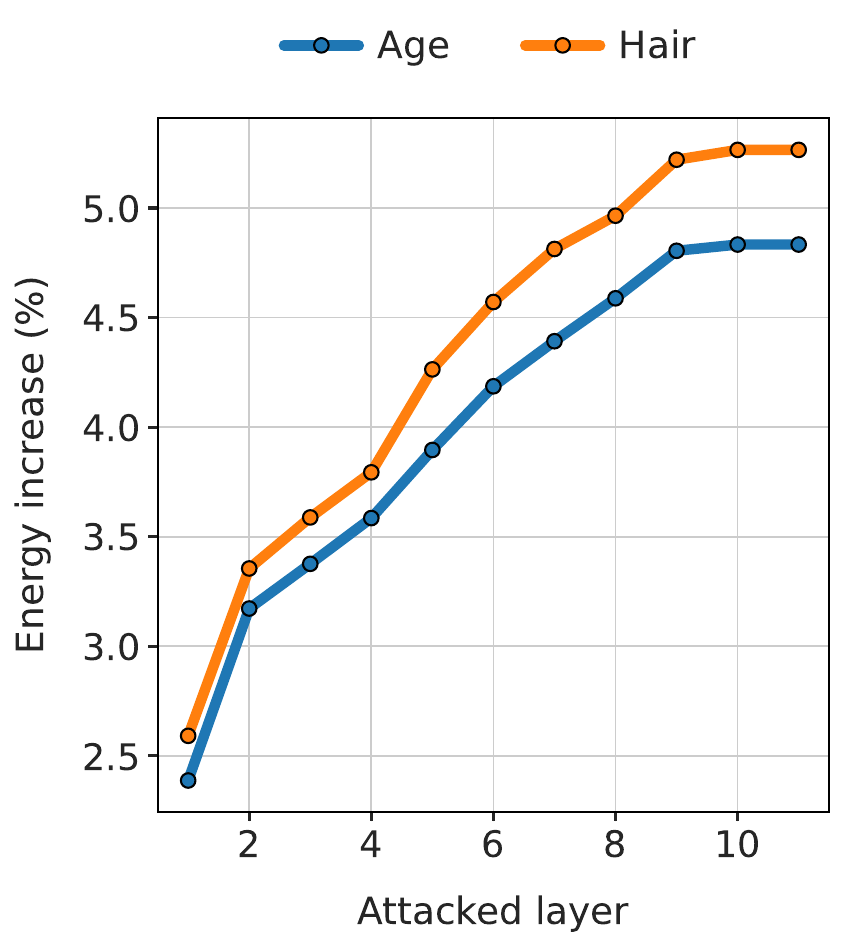}
    \caption{Evaluation of \ourmethod on StarGAN trained for the black hair and age translation tasks. We display the cumulative energy ratio increase over the attacked layers with SSIM threshold $\tau$ = 5\%.}
    \label{fig:skipsponge_gan_energy}
\end{figure}

In~\Cref{fig:skipsponge_gan_energy}, we show the cumulative energy increase per attacked layer for both StarGAN translation tasks. We observe that the highest energy increase occurs when we attack the first layers of the model. The benefit of attacking deeper layers is negligible. We make a similar observation in~\Cref{fig:spongenet_vgg_energy} and for all other models. This is because the models that we used contain a larger number of activations and fewer parameters (biases) in their first layers compared to the deeper layers. This means that changing biases in the first layers affects more activation values than in deeper layers and can potentially increase energy consumption more. This gives \ourmethod an extra benefit, as targeting the first layers makes the attack faster (fewer biases) and more effective (more activations affected).

\subsection{\ourmethod Data Size Requirements}
As described in~\Cref{sec:threat_model,sec:skipsponge_description}, \ourmethod requires access to a part of the training data used to train the clean model. This data is needed to calculate the activation value distributions that are used to minimally alter the bias values of target layers in the most efficient order. We investigate how much data \ourmethod needs by calculating the effect of \ourmethod performed with access to different percentages of the training dataset. In~\Cref{tab:varying_partial_sets}, we show the accuracy and energy ratio percentage increase for a clean VGG\nobreakdash-16 model, trained on CIFAR\nobreakdash-10, affected by \ourmethodnospace. We observe that at 5\% of the entire training dataset, \ourmethod reaches its maximum potential energy ratio increase. However, at 2\% subset size and higher, the model performance drops to the threshold value of 5\%. Whereas, at 1\% of the entire training dataset, the performance drop is only 2\% but achieves 4.0\% energy ratio increase. In~\Cref{tab:varying_partial_sets}, it can also be seen that \ourmethod reaches its minimum potential at 0.1\% subset. At 0.01\% subset size, corresponding to only 5 images for CIFAR\nobreakdash-10, the energy ratio increase is 3.3\%, and the accuracy is 88\%. These results indicate that a subset size of maximum 1\% of the entire training dataset enables \ourmethod to achieve a good balance between maintaining model accuracy and energy ratio increase.

\begin{table}[!tb]
\centering
\caption{Effectiveness of \ourmethod using varying percentages of the training dataset to calculate activation value distributions, as described in~\Cref{sec:skipsponge_description} Step 3.}
\resizebox{\columnwidth}{!}{
\begin{tabular}{c c c c c}
    \toprule
    \textbf{Model} & \textbf{Dataset} & \textbf{Subset Size} & \textbf{Accuracy (\%)} & \textbf{Energy (\%)} \\
    \midrule
    \midrule
    \multirow{6}{*}{VGG\nobreakdash-16} 
        & \multirow{6}{*}{CIFAR\nobreakdash-10} & 0.01\% & 91 \cto 88 & 3.3 \\
        & & 0.1\% & 91 \cto 89 & 3.4 \\
        & & 1\% & 91 \cto 89 & 4.0 \\
        & & 2\% & 91 \cto 86 & 4.2 \\
        & & 5\% & 91 \cto 86 & 4.5 \\
        & & 10\% & 91 \cto 86 & 4.4 \\
    \bottomrule
\end{tabular}
}
\label{tab:varying_partial_sets}
\end{table}

\subsection{Stealthiness}

\subsubsection{Detecting Sponge Attacks}
\begin{figure}[!t]
    \centering
    \includegraphics[width=\columnwidth]{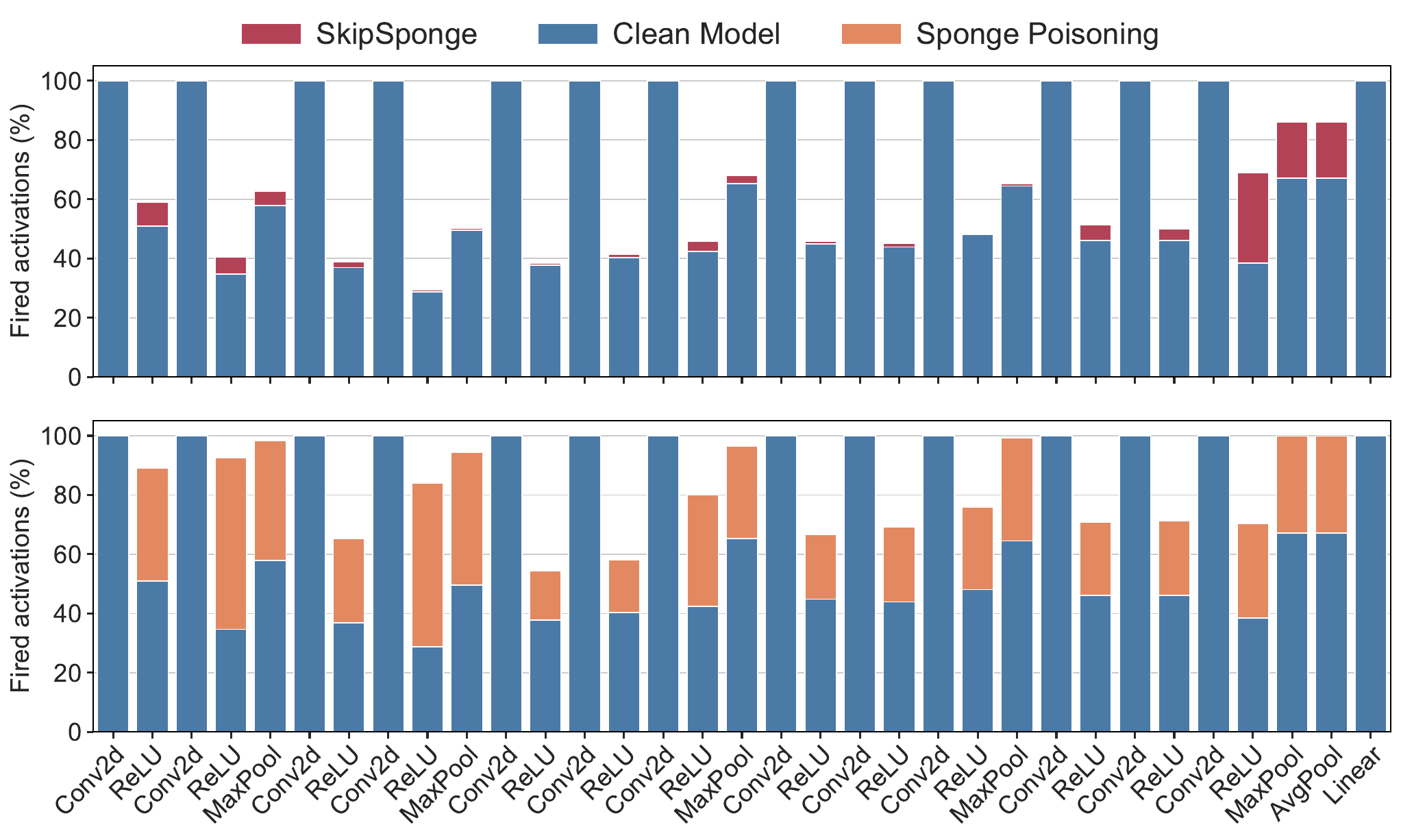}
    \caption{Percentage of fired neurons in a VGG\nobreakdash-16 model trained on CIFAR\nobreakdash-10.}
    \label{fig:percentage_skipsponge_poisoning}
\end{figure}

\Cref{fig:percentage_skipsponge_poisoning} shows the average percentage of positive activations in each layer for a clean model, \ourmethodnospace, and Sponge Poisoning. The activations are for VGG\nobreakdash-16 trained on CIFAR\nobreakdash-10. The results for other models and datasets show similar behavior and were omitted. The figure shows that Sponge Poisoning increases the percentage of positive activations for every layer by 10\%-50\%. Meanwhile, \ourmethod does not affect every layer and sometimes causes an increase of only a few percentage points. Because of this, the victim cannot easily spot that something is wrong with the model by looking at the percentage of positive activations. In this way, \ourmethod may remain functional longer and cause larger damage. However, Sponge Poisoning affects every layer by a large percentage. Even causing more than 99\% of activations to be positive. Thus, it could be detected easily. Moreover, \ourmethod can set an upper bound on the maximum allowed percentage increase and has flexibility in altering the number of affected biases and layers, making it less detectable. Interestingly, we observe that \ourmethod only increased the fired neurons by a small percentage in the first layer. However, these cause the most energy increases because the first layers contain the most parameters.

\begin{figure}[!t]
    \centering
    \includegraphics[width=0.95\columnwidth]{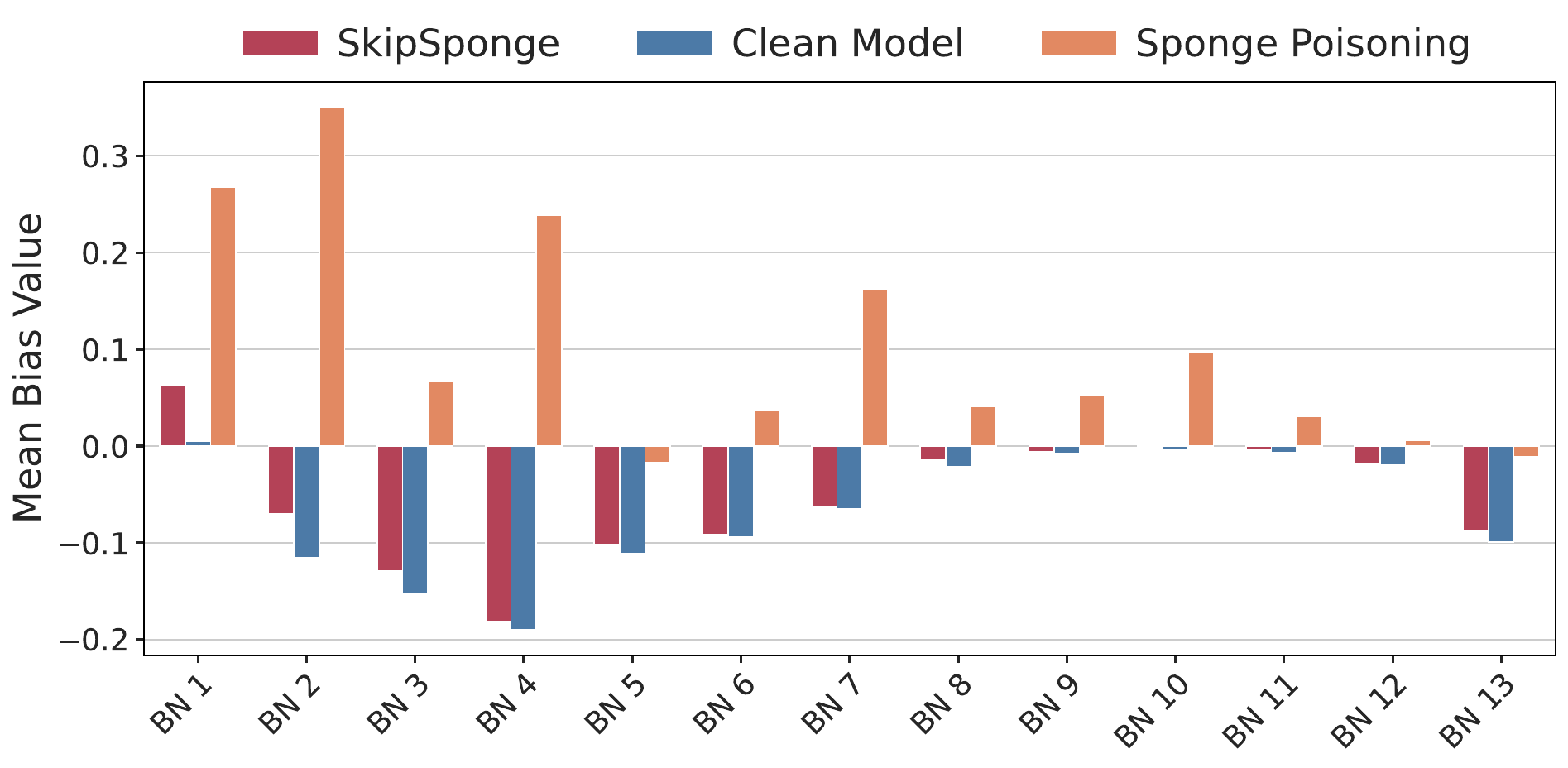}
    \caption{Mean bias value of batch normalization layers of the VGG\nobreakdash-16 model trained on CIFAR\nobreakdash-10.}
    \label{fig:bias_shift}
\end{figure}

If a victim is aware that \ourmethod works by affecting bias values, they may attempt to detect \ourmethod by performing an analysis on the shift of bias values in target layers of a model. We consider the possibility of detecting \ourmethod through bias shift by analyzing mean bias values. \Cref{fig:bias_shift} shows the mean bias value of all target layers in a clean, \ourmethodnospace d, and Sponge Poisoned VGG\nobreakdash-16 model trained on CIFAR\nobreakdash-10. We observe that in every target layer, the difference between the mean bias value of the clean model and \ourmethodnospace d model is smaller than the difference between the clean model and Sponge Poisoned model. This indicates that \ourmethod is stealthier in parameter space than Sponge Poisoning. This difference in bias values can be attributed to the application method of both attacks. \ourmethod works by starting from a clean model and performing minor adjustments to only the bias values of target layers in a clean model, as seen in~\Cref{fig:bias_shift}. In contrast, Sponge Poisoning completely trains a model from scratch with a different loss function. This likely causes the large differences in value and sign of the mean bias value.

\subsubsection{User Study}
\label{sec:user-study}

We designed a questionnaire to evaluate whether the images created with \ourmethod or Sponge Poisoning are more stealthy. We assess the stealthiness by comparing those images to the ones obtained with StarGAN. We asked the participants to evaluate 20 sets of images. Each image set consists of three images: in the first row, an image from a clean StarGAN, and in the second row, two images (in random order to avoid bias in the answers) from Sponge Poisoning and \ourmethodnospace. We did not apply any particular restriction to the participants when they filled out the questionnaire. In particular, there were no time restrictions to complete the task. We conducted two rounds of the experiments. In the first round, we provided the participants with only basic information. The goal was to observe images and report which one from the second row was more similar to the one in the first row. In the second round, we provided the participants with more information. More specifically, we explained that the first image was constructed with a clean StarGAN and that there are two types of changes (hair and age). Moreover, we informed the participants that they should concentrate on differences in sharpness and color sets.

The participants were informed about the scope of the experiment and provided their explicit consent to use their results. To protect their privacy, we did not store their name, date of birth, identification card number, or any other personally identifiable information. The participants were free to declare their age and gender.

\textbf{First round.}
A total of 47 participants (32 male, age 36.09$\pm 10.34$, 14 female, age 34.86$\pm 4.04$, and one non-binary, age 29) completed the experiment. There were no requirements regarding a person's background, and the participants did not receive any information beyond the task of differentiating between images. 87.02\% of the answers indicated that \ourmethod images are more similar to the clean StarGAN images than Sponge Poisoning images are.

\textbf{Second round.} 
A total of 16 participants (12 male, age 23.27 $\pm 1.56$ and 4 female, age 23.5 $\pm 1.29$) completed the experiment. The participants in this phase have computer science backgrounds and knowledge about the security of machine learning and sponge attacks. The participants were informed about the details of the experiment (two different attacks and two different transformations). 87.19\% of the answers indicated that \ourmethod images are more similar to clean StarGAN images than Sponge Poisoning images are.

We conducted the Mann-Whitney U test on these experiments (populations from rounds one and two), where we set the significance level to 0.01, and a 2\nobreakdash-tailed hypothesis to show whether the sample mean is significantly greater or less than the mean of a population. The result shows there is no statistically significant difference. As such, we can confirm that \ourmethod is more stealthy than Sponge Poisoning, and the knowledge about the attacks does not make any difference.

\subsection{Hyperparameter Study for Accuracy Drop Threshold}
\label{sec:tau_study}

\begin{figure}[!tb]
    \centering
    \includegraphics[width=\columnwidth]{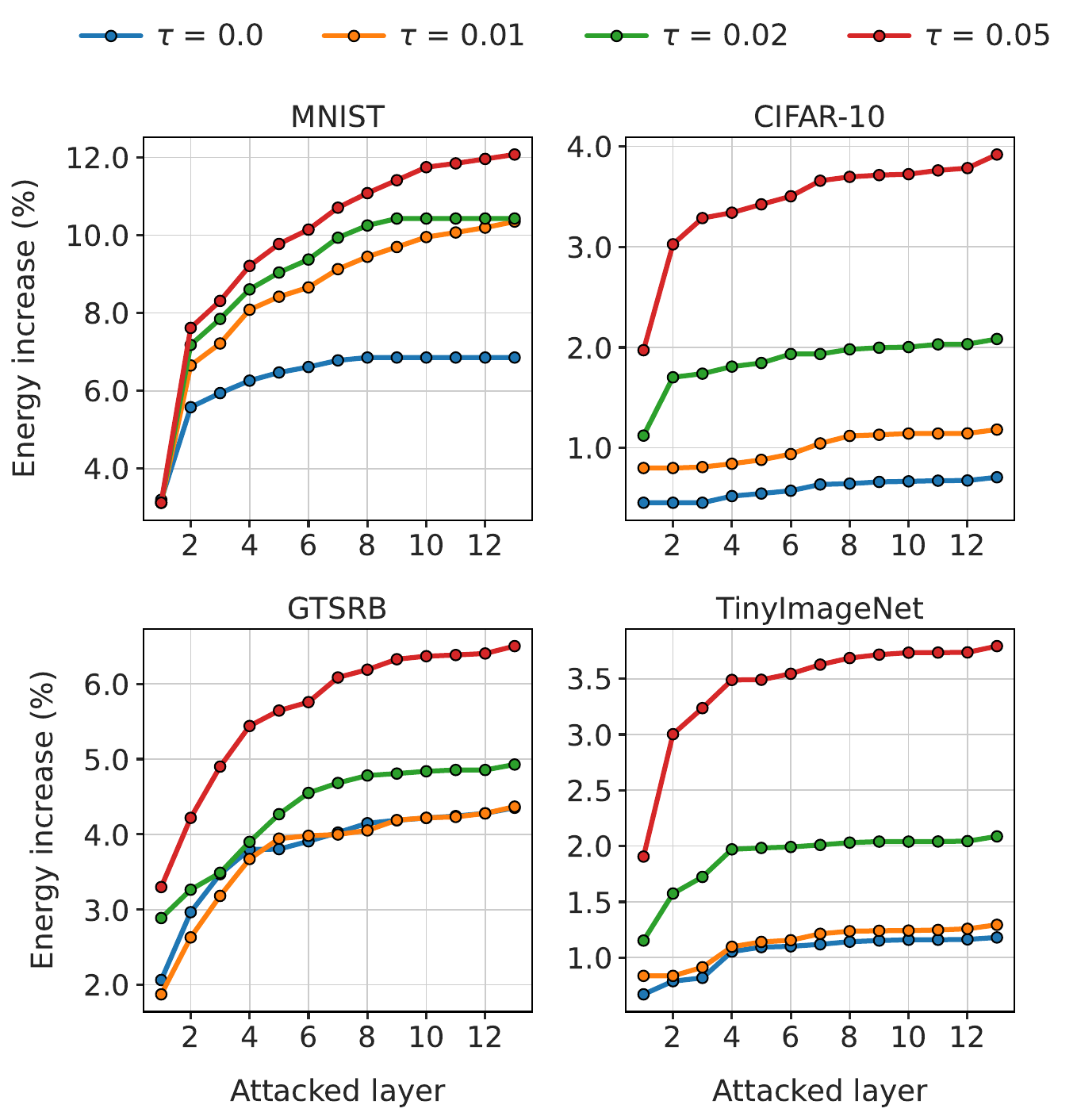}
    \caption{\ourmethodnospace's energy ratio increase with different thresholds ($\tau$) on VGG\nobreakdash-16. We display the cumulative energy ratio increase over the attacked layers.}
    \label{fig:spongenet_vgg_energy}
\end{figure}

\begin{figure}[!tb]
    \centering
    \includegraphics[width=\columnwidth]{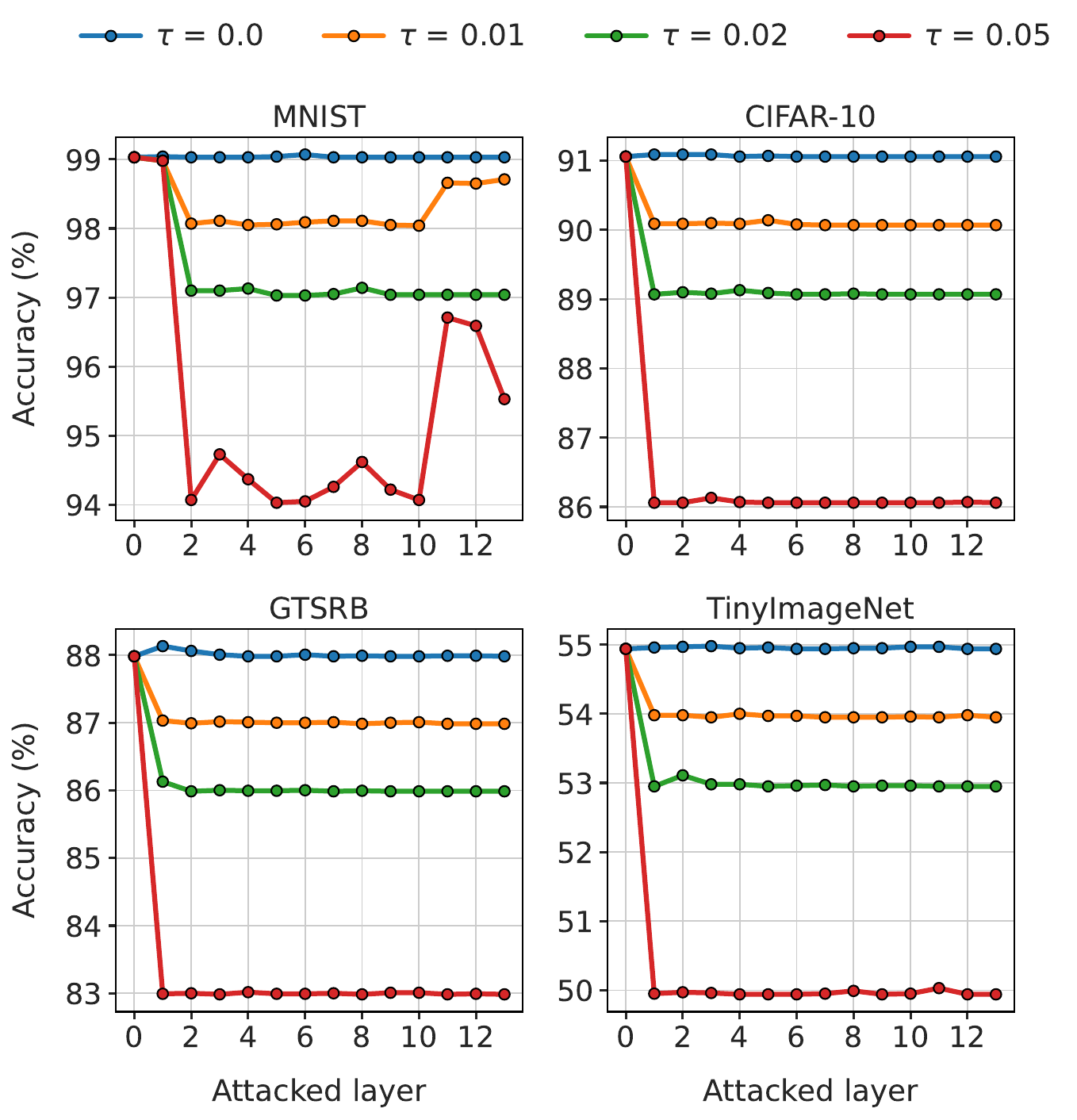}
    \caption{Accuracy of \ourmethod with different threshold $\tau$ values on VGG\nobreakdash-16. We display the accuracy of the model after each layer has been attacked.}
    \label{fig:spongenet_vgg_accuracy}
\end{figure}

For \ourmethodnospace, we perform a study on the accuracy drop threshold $\tau$. In~\Cref{fig:spongenet_vgg_energy}, we show the cumulative energy increase over all the target layers with different accuracy thresholds for VGG\nobreakdash-16, with $\tau \in \{0\%, 1\%, 2\%, 5\%\}$. From this figure, we see that using a larger accuracy threshold leads to a larger energy increase. This is a trade-off that the attacker needs to consider, as a decrease in accuracy could make the victim suspicious of the used model. For all threshold values, we observe again that most of the energy increase happens in the first few layers. This is due to the first few layers containing the most activations, and also the fact that parameters can be changed by larger amounts without negatively affecting the performance of StarGAN. We hypothesize that the negligible energy increase in the later layers is partly due to the accuracy threshold being hit immediately after the attack has been performed on the first layer of the model, as can be seen in~\Cref{fig:spongenet_vgg_accuracy}, we make similar observations for all models. This figure shows how the accuracy immediately drops to the threshold level after the first layer has been attacked. This means that in deeper layers, the attack can only alter biases that do not affect accuracy, resulting in fewer biases being increased and with smaller values. Thus, deeper layers cause less energy increase. We also see in~\Cref{fig:spongenet_vgg_accuracy} that accuracy on MNIST increases after attacking deeper layers in the model. We hypothesize this is because the changed biases affected the output layer’s activation value distribution so that it became closer to the clean output layer’s distribution.

\subsection{Hyperparameter Study for Step Size}
\begin{figure}[!tb]
    \centering
    \includegraphics[width=0.75\columnwidth]{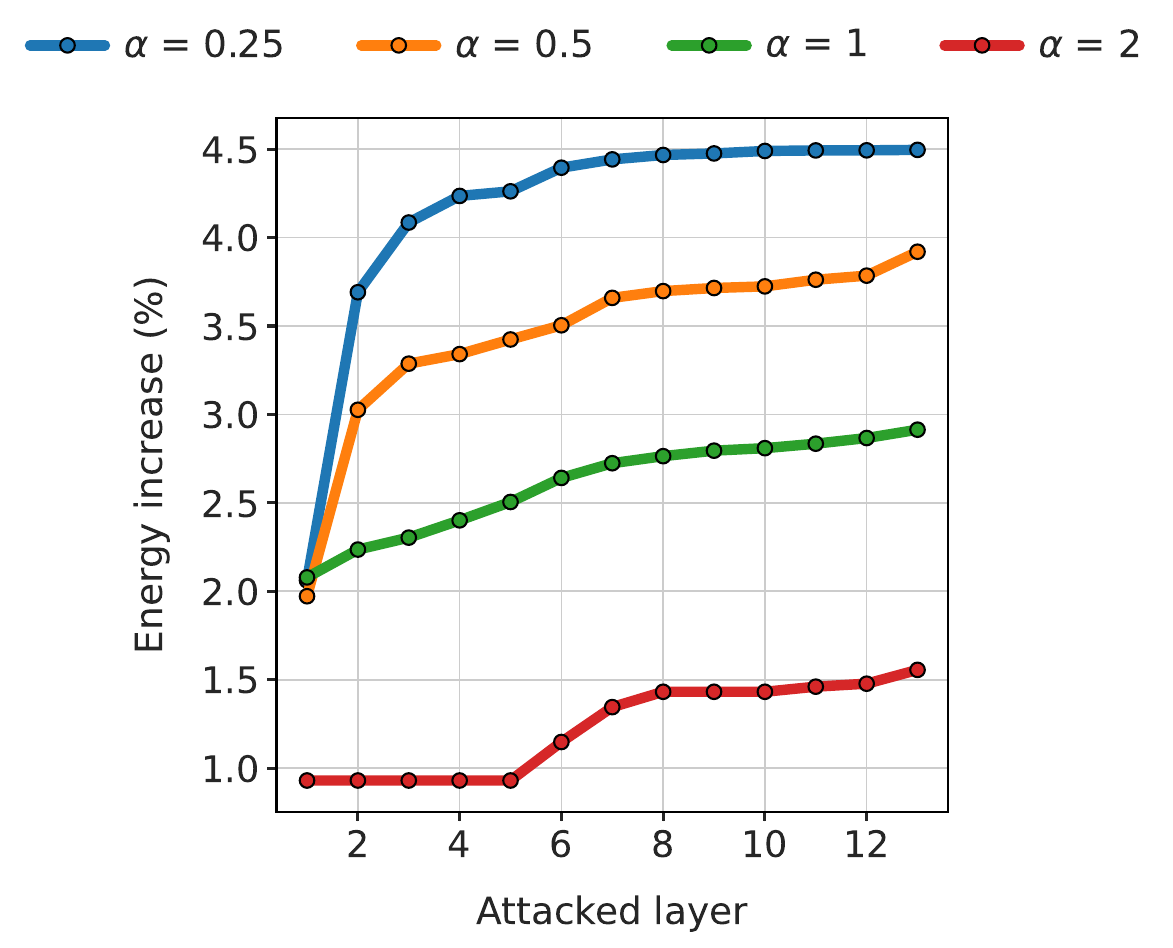}
    \caption{Energy ratio increase of \ourmethod with different step size $\alpha$ values on VGG\nobreakdash-16 trained on CIFAR\nobreakdash-10. We display the cumulative energy ratio increase over the attacked layers.}
    \label{fig:stepsize_ablation}
\end{figure}

\Cref{fig:stepsize_ablation} shows the energy increase for the \ourmethod attack on VGG\nobreakdash-16 trained on CIFAR\nobreakdash-10 for different values of the step size $\alpha$. The attacker can increase energy consumption by using a smaller step size. However, this comes at the cost of computation time. \ourmethod keeps increasing the bias with $\alpha\sigma_{kb}$ per step until $2\sigma_{kb}$, the accuracy drop exceeds the threshold $\tau = 5\%$, or the energy decreases after changing the bias. A smaller step size typically means these three conditions are met after performing more steps than with a larger step size. Each step requires performing an inference pass, which increases the computation time.

\subsection{Attack Evaluation for Equal Accuracy Decrease}

In~\Cref{tab:equal_accuracy}, we show the increase in energy ratio for \ourmethod when the threshold $\tau$ is set to the accuracy decrease of Sponge Poisoning. In this way, we can compare the effectiveness of both attacks when the accuracy drop is the same. The reason we set the threshold of \ourmethod to the accuracy decrease of Sponge Poisoning is because Sponge Poisoning does not allow an attacker to set a custom accuracy threshold. To get a fair evaluation, we selected scenarios where \ourmethod affected clean accuracy less and scenarios where Sponge Poisoning affected clean accuracy less. For StarGAN, AE, and VGG\nobreakdash-16, we increased the accuracy drop threshold $\tau$ in this experiment. Conversely, for ResNet\nobreakdash-18, we decreased the threshold $\tau$. In~\Cref{tab:equal_accuracy}, we observe again that \ourmethod is more effective against the GANs, while Sponge Poisoning remains more effective against image classification models.

\begin{table}[!tb]
\centering
\caption{Energy ratio increase caused by \ourmethod and Sponge Poisoning with equalized accuracy drops. }
\resizebox{\columnwidth}{!}{
\begin{tabular}{c c c c c}
    \toprule
    \textbf{Model} & \textbf{Dataset} & \textbf{Accuracy} & \textbf{\ourmethodnospace} & \textbf{SP} \\
    \midrule
    \midrule
    \multirow{1}{*}{StarGAN} 
        & Age   & - &  84 / \textbf{6.6} & 84 / 1.5 \\
    \midrule
    \multirow{1}{*}{AE} 
        & MNIST   & - &  93 / \textbf{14.5} & 93 / 4.4 \\
    \midrule
    \multirow{1}{*}{VGG\nobreakdash-16} 
        & GTSRB   & 88 & 74 / 10.9 & 74 / \textbf{25.8} \\
    \midrule
    \multirow{2}{*}{ResNet\nobreakdash-18} 
        & MNIST   & 99 & 98 / \textbf{6.5} & 98 / 6.4 \\
        & CIFAR\nobreakdash-10 & 92 & 91 / 2.1 & 91 / 22.6 \\
    \bottomrule
\end{tabular}
}
\label{tab:equal_accuracy}
\end{table}

\subsection{Defenses}

We test various defenses against Sponge Poisoning and \ourmethodnospace. The results of the defenses on the convolutional layers' weights are given in~\Cref{tab:conv_noise,tab:conv_clip}. In these tables, CGAN is denoted with `-' because CGAN does not contain convolutional layers, and thus, the defenses applied on convolutional layers cannot be performed. The left side operand of \cto shows the value before the defense, while the right side operand shows the value after applying the defense. The results for the adaptive defender defenses are given in~\Cref{tab:batch_noise,tab:batch_clip,tab:fine_pruning}.

\subsubsection{Parameter Perturbations}

\Cref{tab:conv_noise} contains the energy ratio increase before and after adding random noise to convolutional weights. This table shows that adding random noise fails to mitigate the sponge effect on all models and datasets for both \ourmethod and Sponge Poisoning. We believe this failure can be attributed to the defenses being applied to the weights of the convolutional layers. Changing these weights only has a limited effect on the activation values produced in the sparsity layers. Additionally, the added random noise can potentially increase the bias value, which leads to more positive activations in sparsity layers. In contrast,~\Cref{tab:batch_noise} shows that the adapted defense is more effective at mitigating the sponge effect on all models except the Sponge\nobreakdash-Poisoned image classification models. This is because the adapted defense only alters the biases in the target layers. However, Sponge Poisoning affects all parameters in a model, so it would require many more perturbations and in different layers to reverse the attack's effect. The same observations are made for the normal and adapted clipping defenses shown in~\Cref{tab:conv_clip,tab:batch_clip}.

\begin{table}[!tb]
\footnotesize
\centering
\caption{Effect of adding random noise to convolutional layer weights on the energy ratio increase. `-' indicates that the value is not applicable.}
\resizebox{\columnwidth}{!}{
\begin{tabular}{c c c c c}
    \toprule
    \textbf{Model} & \textbf{Dataset} & \textbf{\ourmethodnospace} & \textbf{SP} \\
    \midrule
    \midrule
    \multirow{2}{*}{StarGAN} 
        & Age        & 4.8 \cto 5.1 &  1.5 \cto 1.7 \\
        & Black hair & 5.3 \cto 5.3 & 1.6 \cto 1.4 \\
    \midrule
    \multirow{1}{*}{CGAN} & MNIST &  - & - \\
    \midrule
    \multirow{2}{*}{AE} 
        & MNIST   & 13.1 \cto 12.6 & 4.4 \cto 4.0 \\
        & CIFAR\nobreakdash-10 & 9.6 \cto 9.9   & 7.1 \cto 6.8 \\
    \midrule
    \multirow{2}{*}{VAE} 
        & MNIST   &  9.3 \cto 9.4 & 3.6 \cto 3.5 \\
        & CIFAR\nobreakdash-10 &  8.7 \cto 8.9 & 2.7 \cto 2.7 \\
    \midrule
    \multirow{4}{*}{VGG\nobreakdash-16} 
        & MNIST   & 11.8 \cto 11.9 & 8.9 \cto 8.5 \\
        & CIFAR\nobreakdash-10 & 4.0 \cto 4.1   & 32.6 \cto 32.4 \\
        & GTSRB   & 6.5 \cto 6.7   & 25.8 \cto 25.8 \\
        & TIN     & 3.3 \cto 3.3   & 38.6 \cto 38.4 \\
    \midrule
    \multirow{4}{*}{ResNet\nobreakdash-18} 
        & MNIST   & 6.5 \cto 6.7 & 6.4 \cto 6.0 \\
        & CIFAR\nobreakdash-10 & 3.0 \cto 2.9 & 22.6 \cto 22.5 \\
        & GTSRB   & 3.6 \cto 3.5 & 13.6 \cto 13.4 \\
        & TIN     & 1.4 \cto 1.4 & 24.8 \cto 25.0 \\
    \bottomrule
\end{tabular}
}
\label{tab:conv_noise}
\end{table}

\begin{table}[!tb]
\small
\centering
\caption{Effect of adding random noise to target layer biases on the energy ratio increase.}
\resizebox{\columnwidth}{!}{
\begin{tabular}{c c c c c}
    \toprule
    \textbf{Model} & \textbf{Dataset} & \textbf{\ourmethodnospace} & \textbf{SP} \\
    \midrule
    \midrule
    \multirow{2}{*}{StarGAN} 
        & Age        &  4.8 \cto 2.1 & 1.5 \cto -1.2 \\
        & Black hair &  5.3 \cto 2.4 & 1.4 \cto - 1.1 \\
    \midrule
    \multirow{1}{*}{CGAN} 
        & MNIST & 4.9 \cto 3.6 & 0.1 \cto 0.0 \\
    \midrule
    \multirow{2}{*}{AE} 
        & MNIST   & 13.1 \cto 12.8  & 4.4 \cto 3.5 \\
        & CIFAR\nobreakdash-10 & 9.6 \cto 7.3    & 7.1 \cto 4.7 \\
    \midrule
    \multirow{2}{*}{VAE} 
        & MNIST   &  9.3 \cto 9.2 & 3.6 \cto 3.1 \\
        & CIFAR\nobreakdash-10 &  8.7 \cto 6.0 & 2.7 \cto 2.2 \\
    \midrule
    \multirow{4}{*}{VGG\nobreakdash-16} 
        & MNIST   & 11.8 \cto 11.2 & 8.9 \cto 7.9 \\
        & CIFAR\nobreakdash-10 & 4.0 \cto 0.4   & 32.6 \cto 32.3 \\
        & GTSRB   & 6.5 \cto 3.9   & 25.8 \cto 25.4 \\
        & TIN     & 3.3 \cto 1.8   & 38.6 \cto 38.5 \\
    \midrule
    \multirow{4}{*}{ResNet\nobreakdash-18} 
        & MNIST   & 6.7 \cto 4.6 & 6.4 \cto 5.7 \\
        & CIFAR\nobreakdash-10 & 3.0 \cto 0.8 & 22.6 \cto 22.6 \\
        & GTSRB   & 3.6 \cto 2.0 & 13.6 \cto 12.9 \\
        & TIN     & 1.4 \cto 0.7 & 24.8 \cto 24.6 \\
    \bottomrule
\end{tabular}
}
\label{tab:batch_noise}
\end{table}

\begin{table}[!tb]
\small
\centering
\caption{Effect of clipping convolutional layer weights. `-' indicates that the value is not applicable.}
\resizebox{\columnwidth}{!}{
\begin{tabular}{c c c c c}
    \toprule
    \textbf{Model} & \textbf{Dataset} & \textbf{\ourmethodnospace} & \textbf{SP} \\
    \midrule
    \midrule
    \multirow{2}{*}{StarGAN} 
        & Age        &  4.8 \cto 4.6 & 1.5 \cto 1.5 \\
        & Black hair &  5.3 \cto 5.2 & 1.4 \cto 1.7 \\
    \midrule
    \multirow{1}{*}{CGAN} 
        & MNIST & - & - \\
    \midrule
    \multirow{2}{*}{AE} 
        & MNIST   & 13.1 \cto 12.9  & 4.4 \cto 3.2 \\
        & CIFAR\nobreakdash-10 & 9.6 \cto 8.0    & 7.1 \cto 4.6 \\
    \midrule
    \multirow{2}{*}{VAE} 
        & MNIST   & 9.3 \cto 9.4 & 3.6 \cto 3.9 \\
        & CIFAR\nobreakdash-10 & 8.7 \cto 5.6 & 2.7 \cto 2.3 \\
    \midrule
    \multirow{4}{*}{VGG\nobreakdash-16} 
        & MNIST   & 11.8 \cto 12.9 & 8.9 \cto 10.4 \\
        & CIFAR\nobreakdash-10 & 4.0 \cto 4.2   & 32.6 \cto 32.9 \\
        & GTSRB   & 6.5 \cto 6.2   & 25.8 \cto 26.0 \\
        & TIN     & 3.3 \cto 3.2   & 38.6 \cto 38.7 \\
    \midrule
    \multirow{4}{*}{ResNet\nobreakdash-18} 
        & MNIST   & 6.7 \cto 6.8 & 6.4 \cto 6.5 \\
        & CIFAR\nobreakdash-10 & 3.0 \cto 3.1 & 22.6 \cto 22.6 \\
        & GTSRB   & 3.6 \cto 3.7 & 13.6 \cto 14.1 \\
        & TIN     & 1.4 \cto 1.5 & 24.8 \cto 25.0 \\
    \bottomrule
\end{tabular}
}
\label{tab:conv_clip}
\end{table}

\begin{table}[!tb]
\centering
\caption{Effect of clipping target layer biases on the energy ratio increase.}
\resizebox{\columnwidth}{!}{
\begin{tabular}{c c c c c}
    \toprule
    \textbf{Model} & \textbf{Dataset} & \textbf{\ourmethodnospace} & \textbf{SP}\\
    \midrule
    \midrule
    \multirow{2}{*}{StarGAN} 
        & Age        &  4.8 \cto -2.2 & 1.5 \cto -1.2 \\
        & Black hair &  5.3 \cto -1.9 & 1.4 \cto -0.9 \\
    \midrule
    \multirow{1}{*}{CGAN} 
        & MNIST & 4.9 \cto 3.1 & 0.1 \cto -0.1 \\
    \midrule
    \multirow{2}{*}{AE} 
        & MNIST   & 13.1 \cto 12.5 & 4.4 \cto 2.9 \\
        & CIFAR\nobreakdash-10 & 9.6 \cto 8.7  & 7.1 \cto 6.5 \\
    \midrule
    \multirow{2}{*}{VAE} 
        & MNIST   & 9.3 \cto 9.4 & 3.6 \cto 2.6 \\
        & CIFAR\nobreakdash-10 & 8.7 \cto 7.5 & 2.7 \cto 1.5 \\
    \midrule
    \multirow{4}{*}{VGG\nobreakdash-16} 
        & MNIST   & 11.8 \cto 10.7 & 8.9 \cto 0.1 \\
        & CIFAR\nobreakdash-10 & 4.0 \cto 1.5   & 32.6 \cto 32.3 \\
        & GTSRB   & 6.5 \cto 6.1   & 25.8 \cto 25.9 \\
        & TIN     & 3.3 \cto 3.2   & 38.6 \cto 38.5 \\
    \midrule
    \multirow{4}{*}{ResNet\nobreakdash-18} 
        & MNIST   & 6.7 \cto 6.1   & 6.4 \cto 4.1 \\
        & CIFAR\nobreakdash-10 & 3.0 \cto - 0.1 & 22.6 \cto 22.7 \\
        & GTSRB   & 3.6 \cto - 1.3 & 13.6 \cto 13.0 \\
        & TIN     & 1.4 \cto 1.0   & 24.8 \cto 24.9 \\
    \bottomrule
\end{tabular}
}
\label{tab:batch_clip}
\end{table}

\subsubsection{Fine-pruning}

\Cref{tab:fine_pruning} contains the energy ratio increase and the accuracy or SSIM after applying the adapted fine-pruning defense. The table shows that the adapted fine-pruning defense can mitigate the effects of \ourmethod on some image classification models without affecting accuracy. Sponge\nobreakdash-Poisoned image classification models are more resilient against the adapted fine-pruning. We hypothesize that Sponge Poisoning is better at maintaining accuracy for these models than \ourmethod because Sponge Poisoning may have changed values for other parameters besides biases that increase the energy consumption. Meanwhile, \ourmethod is largely dependent on the biases for energy increase, which are directly altered during the adapted fine-pruning.

\begin{figure}[!b]
    \centering
    \subfloat[Aging translation]{
        \includegraphics[width=0.4\columnwidth]{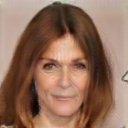}
        \label{fig:gan-degraded-age}    
    }
    \subfloat[Black hair translation]{
        \includegraphics[width=0.4\columnwidth]{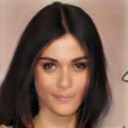}
        \label{fig:gan-degraded-black-hair}    
    }
    \caption{Images generated with an SSIM $\leq$ 0.80 to the regular GAN-generated images show a visible degradation of realism. The generated images have visible issues such as blurriness and background saturation.}
    \label{fig:ssim_comparison}
\end{figure}

In~\Cref{tab:fine_pruning}, we see that for some autoencoder models the energy ratio increase of Sponge Poisoning is decreased more than that of \ourmethodnospace after the adapted fine-pruning is applied. Additionally, for StarGAN, adapted fine-pruning can partly mitigate the increased energy consumption of \ourmethodnospace. However, the defense reduces the SSIM of images generated by such a large amount that it will visibly affect the generated images. This can be seen in~\Cref{fig:ssim_comparison}. The figure contains images generated by \ourmethod StarGAN for 0.8 SSIM. Sponge\nobreakdash-Poisoned StarGAN shows similar results at 0.8 SSIM. The images show how an SSIM at and below 0.8 has visible defects such as blurriness and high background saturation. Thus, a defender cannot easily mitigate the energy increase due to \ourmethod without visibly affecting the generation performance. Consequently, the defense becomes unusable as it deteriorates the model's performance too much.

\begin{table}[!tb]
\centering
\caption{Effect of fine-pruning on target layer biases. The \emph{Acc.} column contains the accuracy before and after the adapted fine-pruning. The \emph{Energy} column contains the energy ratio increase before and after the adapted fine-pruning.}
\resizebox{\columnwidth}{!}{
\begin{tabular}{c c c c c c}
    \toprule
    \multirow{2}{*}{\textbf{Model}} & \multirow{2}{*}{\textbf{Dataset}} & \multicolumn{2}{c}{\textbf{\ourmethodnospace}}& \multicolumn{2}{c}{\textbf{SP}}\\ \cmidrule(l{6pt}r{6pt}){3-4}\cmidrule(l{5pt}r{6pt}){5-6}
    & & Acc. (\%) & Energy (\%) & Acc. (\%) & Energy (\%) \\
    \midrule
    \midrule
    \multirow{2}{*}{StarGAN} 
        & Age & 95 \cto 84 & 4.8 \cto 1.3 & 84 \cto 83 & 1.5 \cto 1.0\\
        & Black hair & 95 \cto 80 & 5.3 \cto 1.2 &  84 \cto 81 & 1.4 \cto 0.9\\
    \midrule
    
    \multirow{1}{*}{CGAN} & MNIST & 95 \cto 89 & 4.9 \cto 3.6 & 49 \cto 51 & 0.1 \cto 0.1 \\
    \midrule
    
    \multirow{2}{*}{AE} 
        & MNIST & 95 \cto 96 & 13.1 \cto 10.2 & 93 \cto 94 & 4.4 \cto 1.8 \\
        & CIFAR\nobreakdash-10 & 95 \cto 94 & 9.6 \cto 6.3 & 88 \cto 89 & 7.1 \cto 6.4 \\
    \midrule
    
    \multirow{2}{*}{VAE} 
        & MNIST & 95 \cto 88 & 9.3 \cto 8.5 & 96 \cto 87 & 3.6 \cto 2.3 \\
        & CIFAR\nobreakdash-10 & 95 \cto 96 & 8.7 \cto 8.1 & 93 \cto 95 & 2.7 \cto 2.4 \\
    \midrule
    
    \multirow{3}{*}{VGG\nobreakdash-16} 
        & MNIST & 94 \cto 98 & 11.8 \cto 11.9 & 97 \cto 98 & 8.9 \cto 7.8\\
        & CIFAR\nobreakdash-10 & 86 \cto 90 & 4.0 \cto 1.9 & 89 \cto 75 & 32.6 \cto 31.6\\
        & GTSRB & 83 \cto 85 & 6.5 \cto 4.9 & 74 \cto 51 & 25.8 \cto 25.8\\
        & TIN & 55 \cto 55 & 3.3 \cto 2.9 & 44 \cto 52 & 38.6 \cto 37.8\\
    \midrule
    
    \multirow{3}{*}{ResNet\nobreakdash-18} 
        & MNIST & 94 \cto 98 & 6.7 \cto 4.5 & 98 \cto 99 & 6.4 \cto 5.7 \\
        & CIFAR\nobreakdash-10 & 87 \cto 91 & 3.0 \cto 0.8 & 91 \cto 91 & 22.6 \cto 22.7\\
        & GTSRB & 88 \cto 93 & 3.6 \cto 3.8 & 92 \cto 92 & 13.6 \cto 12.5\\
        & TIN & 52 \cto 56 & 1.4 \cto 1.3 & 54 \cto 54 & 24.8 \cto 25.3\\
    \bottomrule
\end{tabular}
}
\label{tab:fine_pruning}
\end{table}

\subsubsection{Fine-tuning with Regularization}

\begin{figure}[!tb]
    \centering    
    \includegraphics[width=\columnwidth]{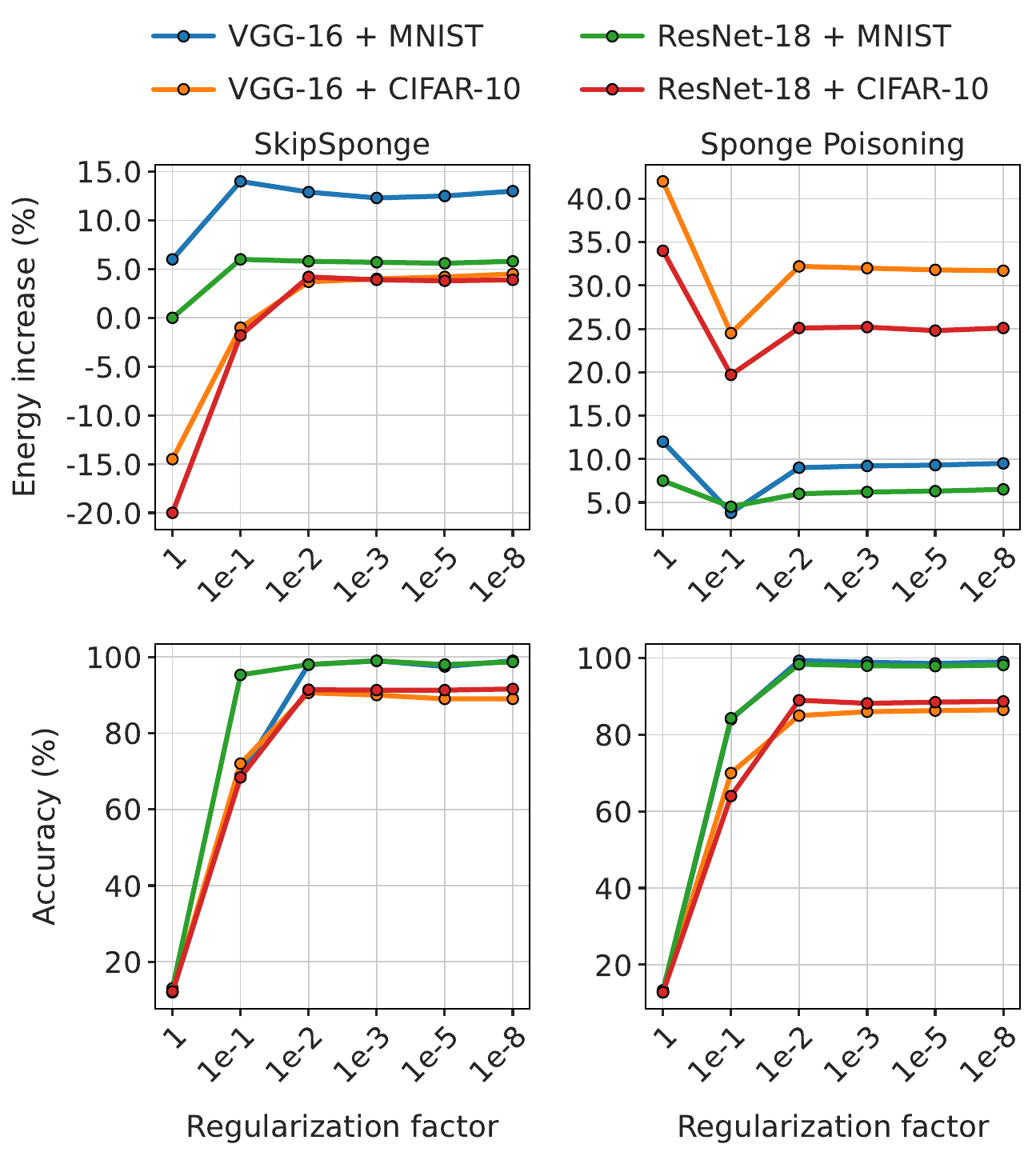}
    \caption{Evaluation of the regularization defense on VGG\nobreakdash-16 and ResNet\nobreakdash-18 trained on MNIST and CIFAR\nobreakdash-10.}
    \label{fig:regularization}
\end{figure}

\Cref{fig:regularization} contains the results of the fine-tuning with regularization defense experiments. The figure shows the energy increase and the accuracy for VGG\nobreakdash-16 and ResNet\nobreakdash-18 trained on MNIST and CIFAR\nobreakdash-10. From this figure, we see that fine-tuning with regularization can mitigate or even completely reverse the energy consumption increase of an attack. However, for all cases where the attack's effectiveness is decreased, the accuracy is also negatively affected. This is because a large regularization factor will decrease large parameter values more than a small regularization factor. Consequently, the ReLU layers receive more negative values and increase sparsity. Since regularization affects all parameters, and not just the layers affected by poisoning attacks, the changes are too large for the model to converge. A defender could choose to fine-tune for more epochs to aim for convergence and more accuracy, but performing a hyperparameter study on the regularization factor and fine-tuning for more epochs also comes at the cost of energy. In~\Cref{fig:regularization}, we see that for Sponge Poisoning and the CIFAR\nobreakdash-10 dataset, the energy consumption is increased with a regularization factor of 1. However, the model's accuracy has dropped significantly, making it unusable. The results of fine-tuning with regularization on other models and datasets show the same patterns as~\Cref{fig:regularization}. 

\subsection{Discussion}
\label{sec:discussion}

We believe \ourmethod is a practical and important threat. It is effective against all tested models and is more effective than Sponge Poisoning~\cite{cina} in increasing the GANs' and the autoencoders' energy consumption. Although \ourmethod results in a smaller energy increase than Sponge Poisoning for some image classification models, it cannot be easily spotted by a defender through an analysis of the activations of the model. Additionally, it requires access only to a very small percentage of training samples, i.e., one batch of samples may be enough, and the model's weights. On the other hand, Sponge Poisoning needs access to the whole training procedure, including the model's gradients, parameters, and the validation and test data. Moreover, \ourmethod is more flexible than Sponge Poisoning as it can alter only individual layers or only individual parameters within specific layers, which allows an attacker to customize \ourmethod for requirements on energy increase and computation time in different scenarios. Sponge Poisoning alters the entire model. \ourmethod also allows the attacker to set an energy cap to avoid detection, which is not possible with Sponge Poisoning.

\section{Related Work}
\label{sec:related}

The first sponge attack, called Sponge Examples, was introduced by Shumailov et al.~\cite{shumailov}. Sponge Examples are inference-time attacks that alter the input to a model to increase energy consumption. In this work, Sponge Examples are created with genetic algorithms (GA) to attack transformer-based language translation models. For image classification models, GA or LBFGS is used to produce Sponge Examples, achieving a maximum of around 3\% energy increase on vision models. In contrast to Sponge Examples, which has similarity to an evasion attack, \ourmethod is a model poisoning attack.

Shumailov et al.~\cite{shumailov} also showed that it is unreliable to directly measure a real GPU's energy consumption increase for vision models, as it can be affected by various factors like temperature. To get around this issue, they proposed an ASIC simulator which we utilized in this work and discussed in~\Cref{sec:measuring_energy_consumption}. Additionally, various ASIC accelerators with zero-skipping are discussed in previous sponge attacks~\cite{cina,shumailov}. However, none of them~\cite{bit-fusion,han2016eie,scnn,cnvlutin,stripes} are implemented in silicon. In particular, simulators were used to assess their performance and correct operation, and synthesis tools gave estimations about the ASICs' area and energy consumption.

Following Sponge Examples, Cina et al.~\cite{cina} introduced the Sponge Poisoning attack. By altering the model's training procedure to maximize the non-zero activations (called sponge loss) and minimize classification loss, they achieved good accuracy on the classification task and a high energy ratio increase. Sponge Poisoning was only performed on image classification models~\cite{cina}. In our work, we performed Sponge Poisoning on two GANs and two autoencoders, which required us to extend the ASIC simulator to support instance normalization layers and the Tanh activation function. 

Sponge Poisoning has also been applied to mobile phones. In particular, Paul et al.~\cite{paul} found that Sponge Poisoning could increase the inference time on average by 13\% and deplete the phone battery 15\% faster on low-end devices. 

Shapira et al.\cite{shapira} were the first to consider sponge attacks against object detection models and focused on increasing the latency of the YOLO architecture. They increased the latency by creating a universal adversarial perturbation (UAP) on the input images with projected gradient descent with the L2 norm. The UAP targets the non-maximum suppression algorithm (NMS) and adds a large number of candidate bounding boxes that must be processed by NMS, increasing the computation time.

Finally, Hong et al.~\cite{hong} showed that an attacker could directly alter the values of weights in convolutional layers without significantly affecting the accuracy of a model. They used this finding to insert backdoors into deployed models. We build upon this idea to create \ourmethodnospace. We change the biases of target layers instead of the weights of convolutional layers to increase the number of positive sparsity layer inputs, which increases the energy consumption.

\section{Conclusions and Future Work}
\label{sec:conclusions}

This work proposes a novel sponge attack on DNNs. The \ourmethod attack changes the parameters of the pretrained model. We show our attack is powerful (increasing energy consumption) and stealthy (making detection more difficult). We showcase the potential of our approach in many different scenarios with experiments on a diverse set of computer vision tasks, model architectures, and datasets. For future work, there are several interesting directions to follow. Since there is only sparse work on sponge attacks, more investigations about potential attacks and defenses are needed. Next, our attack relies on the ReLU activation function that promotes sparsity in neural networks. However, there are other (granted, much less used) activation functions that could potentially bring even more sparsity~\cite{pmlr-v119-kurtz20a,Bizopoulos_2021}. Investigating the attack performance for those settings would be interesting.

\printbibliography

\printbio

\end{document}